\documentclass[runningheads]{llncs}

\usepackage{amssymb}
\usepackage{tabularx}
\usepackage{siunitx}
\usepackage{makecell}
\usepackage{tikz}
\usetikzlibrary{arrows.meta, positioning, shapes.geometric, calc, shadows}
\usepackage{pgf-umlsd}
\usepackage{hyperref}
\usepackage{subcaption}
\usepackage{csquotes}
\usepackage{booktabs}

\usepackage[T1]{fontenc}

\usepackage{graphicx}

\usepackage{color}

\usepackage{wrapfig}

\makeatletter
\long\def\@makecaption#1#2{%
  \vskip\abovecaptionskip
  \sbox\@tempboxa{\footnotesize #1. #2}%
  \ifdim \wd\@tempboxa >\hsize
    \footnotesize #1. #2\par
  \else
    \global \@minipagefalse
    \hb@xt@\hsize{\hfil\box\@tempboxa\hfil}%
  \fi
  \vskip\belowcaptionskip}
\makeatother

\newcommand{\Subsubsection}[1]{\subsubsection{#1.}}

\begin{document}

\title{From Bracha to Coded MBRB: Benchmarking Byzantine Reliable Broadcast Implementations\thanks{This technical report complements the conference version of this work, which will appear in the proceedings of SSS 2026.
}}
\titlerunning{Reliable Broadcast Benchmarking}

\author{
Yenan Wang\inst{1}\orcidID{0009-0004-1389-1196} \and
Jesper Kullberg\inst{1} \and
Fabian Paglianno Persson\inst{1} \and
Elad Michael Schiller\inst{1}\orcidID{0000-0003-3258-3696} \and
Timoth\'e Albouy\inst{2}\orcidID{0000-0001-9419-6646}
}

\authorrunning{Y. Wang et al.}

\institute{
Chalmers University of Technology, Gothenburg, Sweden\\
\email{
\{yenan,jesperku,fabianpa,elad\}@chalmers.se
}
\and
IMDEA Software Institute, Madrid, Spain\\
\email{timothe.albouy@imdea.org}
}
\maketitle
\begin{abstract}
Byzantine Reliable Broadcast (BRB) and Message-Adversary-Tolerant Byzantine Reliable Broadcast (MBRB) are reliable-dissemination abstractions for fault-tolerant distributed systems. Yet their operational behavior is shaped not only by specifications and asymptotic communication bounds, but also by serialization, cryptography, buffering, orchestration, deployment environment, and fault-injection semantics. This paper implements and evaluates Bracha [Information and Computation, 1987], AFRT by Albouy et al. [TCS, 2023], and Coded MBRB by Albouy et al. [OPODIS, 2024]. We implement the algorithms in a shared Go codebase with common orchestration, instrumentation, parser-based specification checks, fault injection, and an open-source reproducibility artifact. The evaluation uses single-shot broadcasts in the Shadow network simulator, native profiling, a Google Cloud Platform deployment, and a distributed FABRIC testbed, covering controlled experiments up to 30 nodes, payloads up to 40 MB, 92,190 runs, and 2,361,600 parser-checked entries. The results show that Coded MBRB reduces transmitted data and improves latency in the evaluated cloud setting for larger payloads, but shifts cost to cryptographic and coding computation. Bracha and AFRT incur lower CPU costs at smaller payloads, but their full-payload dissemination increases processing, allocation, and network costs as payloads grow. Across the tested configurations, the parser found no duplicate deliveries, conflicting deliveries, or deliveries of values different from the sender's payload. The paper contributes implementation-level evidence and an extensible artifact for benchmarking BRB and MBRB as executable distributed-system components, exposing bottlenecks and operational trade-offs that are hidden by algorithmic descriptions alone.

\end{abstract}

\section{Introduction}

\label{sec:introduction}
Reliable broadcast (RB) is a fundamental communication abstraction at the center of many fault-tolerant distributed systems.
RB lets a designated sender disseminate an application value $m$ so that correct (non-faulty) nodes deliver at most one common value; and if the sender is correct, all correct nodes deliver $m$. RB plays a crucial role in many practical applications, such as event notification, state machine replication~\cite{DBLP:conf/podc/AbrahamN0X21}, or asset transfer~\cite{DBLP:journals/eatcs/AuvolatFRT20,DBLP:conf/dsn/CollinsGKKMPPST20}. Fault tolerance comes in multiple flavors: if faulty nodes can behave arbitrarily, we talk about Byzantine Reliable Broadcast (BRB); and if, in addition to Byzantine failures, a message adversary (MA) can suppress implementation messages exchanged between correct nodes, we talk about MA-Tolerant Byzantine Reliable Broadcast (MBRB). MBRB preserves safety but replaces the all-node termination of BRB with a quantified delivery-power guarantee.
In an asynchronous message-passing system of $n$ nodes, where up to $t$ can be Byzantine and an MA that can omit up to $d$ copies of each algorithm-generated broadcast to correct recipients, it has been shown that BRB (which assumes $d=0$) can be solved if and only if $n>3t$~\cite{DBLP:journals/iandc/Bracha87,DBLP:books/sp/Raynal18}, and MBRB can be solved if and only if $n>3t+2d$~\cite{DBLP:journals/tcs/AlbouyFRT23}.

Rather than proposing a new algorithm, we compare how implementations of these BRB and MBRB algorithms behave as executable systems, where software choices expose costs hidden by abstract algorithm descriptions.

\Subsubsection{Algorithms and scope}
{We implement Bracha's BRB~\cite{DBLP:journals/iandc/Bracha87,DBLP:books/sp/Raynal18}, AFRT~\cite{DBLP:journals/tcs/AlbouyFRT23}, and Coded MBRB~\cite{DBLP:conf/opodis/AlbouyFGHRSTZ24}. Bracha is the classical BRB baseline, whereas AFRT and Coded MBRB both implement MBRB.
In particular, Bracha assumes reliable channels (\(d=0\)) and tolerates Byzantine nodes, whereas the MBRB algorithms additionally tolerate bounded omissions of implementation messages, and Coded MBRB uses erasure-coded fragments and cryptographic evidence to reduce the resulting data movement.
Thus, Bracha provides a reference for the cost of adding message-adversary tolerance, while AFRT versus Coded MBRB is the direct like-for-like comparison. AFRT forwards full payloads with accumulated signatures, whereas Coded MBRB uses erasure coding and cryptographic evidence to reduce payload movement.}
Recent communication-efficient BRB work shows that the design space remains active~\cite{DBLP:conf/opodis/Locher24,DBLP:conf/opodis/Locher25,DBLP:conf/eurocrypt/LocherS25}; we therefore treat the studied algorithms as representative implementation points focused on BRB and MBRB under Byzantine faults and message-adversary assumptions, not as exhaustive coverage of all BRB designs. {Recent coded BRB algorithms address communication efficiency under the classical reliable-channel model~\cite{DBLP:conf/podc/AlhaddadDD0VXZ22,DBLP:conf/ccs/DasX021}; our direct comparison instead focuses on AFRT and Coded MBRB, which implement the same MBRB abstraction.}

Our implementations support single-shot broadcast instances. This choice isolates the cost of one invocation of the broadcast primitive and avoids confounding the measurements with batching, sequence-number management, sliding windows, concurrent instances, admission control, or garbage collection across repeated broadcasts. Consequently, the paper characterizes per-broadcast communication, CPU, memory, latency, and delivery behavior, but does not measure sustained throughput of a long-running dissemination service.

\Subsubsection{Approach}
The algorithms are implemented in a shared Go codebase as peer-to-peer nodes coordinated by a centralized experiment controller~\cite{brb-eval-open-source-artifact}. The controller distributes configuration, peer lists, cryptographic material, payload size, selected algorithm, and fault-injection roles, and acts as a synchronization point for registration, readiness, execution, completion, and metric collection. Nodes use persistent pairwise TCP connections and a manager layer that records message and byte counters, timestamps delivery, routes incoming messages to the selected algorithm, and provides the common interception point for fault injection. The implementation uses established libraries for erasure coding and cryptographic operations~\cite{gnark-crypto-v0.20.1,klauspost-reedsolomon}.

The evaluation combines Shadow, which is a discrete-event network simulator, native profiling, {Google Cloud Platform (GCP)} deployment, and supplementary FABRIC measurements. Shadow executes the same Go binaries over a deterministic simulated topology for communication-cost and fault-injection experiments~\cite{DBLP:conf/usenix/JansenNW22,shadow_guide}. Native profiling measures user-space CPU instructions with Linux \texttt{perf} and Go runtime memory statistics. GCP provides the primary deployment-oriented latency measurements using controller-measured timing over private VPC communication, while FABRIC provides supplementary latency evidence over a more distributed infrastructure. Fault injection covers silent nodes, randomized omission of implementation messages, and one partition-based equivocating-sender scenario inspired by Twins~\cite{DBLP:conf/opodis/BanoSCPLCM21}. Parser checks detect duplicate delivery, conflicting delivery, invalid delivery, and spurious delivery; these consistency tests provide implementation-level evidence, not exhaustive Byzantine testing.

\Subsubsection{Findings}
The experiments expose a concrete systems trade-off. In Shadow at $n=30$ with a 1~MB payload, Coded MBRB transmits approximately 128~MB, whereas Bracha and AFRT transmit about 1.7~GB, giving roughly a 14$\times$ reduction in transmitted data. Under payload scaling at $n=30$, Bracha and AFRT reach about 14~GB at an 8~MB payload, while Coded MBRB remains around 1~GB. This communication advantage is not free: in native profiling at $n=30$ with a 1~MB payload, Coded MBRB executes about 800 million CPU instructions per node, compared with roughly 250 million for Bracha and AFRT. However, this computational disadvantage reverses at larger payloads, with Bracha overtaking Coded MBRB in CPU cost around 6~MB. Full-payload dissemination also creates substantial memory pressure: at an 8~MB payload, Bracha and AFRT exceed 800~MB peak heap per node, while Coded MBRB remains around 350~MB. In the GCP deployment at $n=5$ and an 8~MB payload, Bracha and AFRT reach about 900~ms completion latency, while Coded MBRB remains below 400~ms. Across 92,190 runs and 2,361,600 parser-checked entries, the parser found no duplicate delivery, conflicting delivery, invalid delivery, or spurious delivery in the tested configurations. These results provide reproducible implementation-level evidence about selected operating regimes.

\Subsubsection{Our contributions}
This paper contributes the following.
\begin{enumerate}
\item We present an implementation architecture for evaluating Bracha's BRB, AFRT, and Coded MBRB algorithms as executable distributed systems, including reusable orchestration, transport, message-management, metric-collection, and fault-injection components.
\item We provide an open-source reproducibility artifact, including code, configurations, automation scripts, raw-output processing, parser-based consistency tests, and plotting support~\cite{brb-eval-open-source-artifact}.
\item We conduct an empirical study of Bracha's BRB, AFRT, and Coded MBRB across simulation, native profiling, and deployments on GCP and FABRIC, characterizing communication cost, local computation, memory behavior, latency, delivery behavior, and consistency-check outcomes under controlled parameter scaling and selected injected faults.
\end{enumerate}

\noindent To facilitate reproducibility, our implementation and artifacts can be found in our open-source code repository~\cite{brb-eval-open-source-artifact}.

\section{Related Work}
\label{sec:extended-related-work}

In this section, we present and expand the compact related-work positioning given in the introduction. The goal is not to provide a complete survey of reliable broadcast, but to clarify how the present implementation-driven study relates to algorithmic BRB and MBRB work, communication-efficient broadcast, implementation artifacts, larger BFT systems, and benchmarking or testing methodology.

\subsection{Reliable-Broadcast Abstractions and Algorithmic Variants}
\label{sec:rw-abstractions}

Byzantine reliable broadcast originates from the need to disseminate a sender value in the presence of Byzantine behavior while preserving agreement-like delivery properties among correct processes. Bracha's asynchronous construction established the echo/ready structure that remains a standard reference point for fully connected asynchronous systems~\cite{DBLP:journals/iandc/Bracha87}. Raynal's treatment places Byzantine reliable broadcast among the basic communication abstractions of fault-tolerant message-passing systems and provides the formulation used as the classical baseline in this paper~\cite{DBLP:books/sp/Raynal18}. These works define the baseline abstraction but do not by themselves answer how executable implementations behave under concrete serialization, buffering, cryptographic, and deployment choices.

Several later directions broaden the model. Dolev's work on reliable communication in unknown and unreliable environments underlies reliable dissemination over non-complete networks~\cite{DBLP:conf/focs/Dolev81}. Practical Byzantine reliable broadcast on partially connected networks combines Bracha-style broadcast with Dolev-style communication and studies optimizations for graph connectivity and path diversity~\cite{DBLP:journals/corr/abs-2104-03673}. Other directions modify the delivery semantics or execution setting. Byzantine-tolerant causal broadcast layers causal delivery over Byzantine-tolerant dissemination~\cite{DBLP:journals/tcs/AuvolatFRT21}. Set-constrained delivery broadcast constrains the sets of values that can be delivered together~\cite{DBLP:conf/opodis/AuvolatRT19}. Repeated and amortized broadcast work studies how repeated invocations, clients, or long-running services can change the average cost of broadcast~\cite{DBLP:conf/wdag/CamaioniGMV22,DBLP:journals/corr/abs-2209-13304,DBLP:conf/podc/WanM0SX23,DBLP:journals/tcs/DuvignauRS23}.

These works define a broad algorithmic design space. Our study does not try to cover all reliable-broadcast abstractions, topologies, or repeated-execution variants. Instead, it uses Bracha, AFRT, and Coded MBRB as three representative points that expose different implementation regimes: classical full-payload BRB, signature-based message-adversary-tolerant broadcast, and coded message-adversary-tolerant broadcast.
Recent work on communication-efficient BRB shows that the design space remains active, including algorithms with low communication and time complexity~\cite{DBLP:conf/opodis/Locher24}, reduced cost in failed executions~\cite{DBLP:conf/opodis/Locher25}, and MiniCast-style long-message communication complexity~\cite{DBLP:conf/eurocrypt/LocherS25}. This broader landscape motivates treating our three implementations as representative points focused on BRB and MBRB under Byzantine faults and message-adversary assumptions, rather than as exhaustive coverage of all reliable-broadcast designs.

\subsection{Message-Adversary-Tolerant and Communication-Efficient Broadcast}
\label{sec:rw-communication-efficient}

The AFRT paper introduces message-adversary-tolerant Byzantine reliable broadcast (MBRB), where a network-level adversary may omit selected implementation messages and the abstraction is weakened from delivery by all correct processes to quantified delivery power~\cite{DBLP:journals/tcs/AlbouyFRT23}. The corresponding condition $n>3t+2d$ makes both Byzantine faults and message-omission power explicit. Coded MBRB refines this line by targeting near-optimal communication under the message-adversary model, replacing repeated full-payload forwarding with erasure-coded fragments, vector commitments, and threshold signatures~\cite{DBLP:conf/opodis/AlbouyFGHRSTZ24}. These two algorithms are central to our evaluation because they make a direct systems trade-off visible: lower communication can require additional cryptographic and coding computation.
{We refer the interested reader to~\cite{DBLP:phd/hal/Albouy24} for a monograph on the MBRB problem.}

{Classical asynchronous BRB has also seen substantial progress in coded
communication-efficient designs. Das, Xiang, and Ren~\cite{DBLP:conf/ccs/DasX021} present an
asynchronous data dissemination primitive that yields communication-efficient
Byzantine reliable broadcast, while Alhaddad et al.~\cite{DBLP:conf/podc/AlhaddadDD0VXZ22} propose a balanced
BRB protocol with near-optimal communication and improved computation.
Unlike AFRT and Coded MBRB, these algorithms assume the classical
reliable-channel model rather than the MBRB message-adversary model.
Consequently, they complement our study rather than serving as direct
algorithmic baselines for the AFRT versus Coded MBRB comparison.}

{Beyond these coded BRB designs, recent work on communication-efficient Byzantine reliable broadcast further emphasizes that communication cost remains an active algorithmic concern.} Locher's work on low communication and time complexity studies asynchronous Byzantine reliable broadcast and reduces the overhead factor of coded reliable-broadcast designs under appropriate execution conditions~\cite{DBLP:conf/opodis/Locher24}. Follow-up work on the failure case introduces reliable-broadcast detectors and studies how to reduce communication cost when the sender fails or no value is delivered~\cite{DBLP:conf/opodis/Locher25}. MiniCast minimizes long-message communication complexity for reliable broadcast, and later work aims to improve its round complexity~\cite{DBLP:conf/eurocrypt/LocherS25,DBLP:journals/iacr/LocherS25}. These works are important for positioning because they show that the selected algorithms are not the final word on communication efficiency.

The present paper is complementary to these algorithmic advances. We do not claim that Coded MBRB is the most communication-efficient BRB design known today, nor do we evaluate all recent low-communication BRB algorithms. Rather, we use Coded MBRB as a recent message-adversary-tolerant coded design whose implementation exposes the concrete cost of replacing full-payload dissemination with cryptographic and coding work. Adding Locher-style and MiniCast-style algorithms to the same harness is a natural next step enabled by the artifact.

\subsection{Reliable-Broadcast Implementations and Artifacts}
\label{sec:rw-implementations}

Implementation-oriented reliable-broadcast work is more limited than the algorithmic literature. Practical Byzantine reliable broadcast on partially connected networks reports a C++ implementation and evaluates optimized Bracha-Dolev combinations using a real C++ implementation and actual deployment~\cite{DBLP:journals/corr/abs-2104-03673}. Reliable Broadcast in Practical Networks provides a Go-based framework using Mininet to evaluate reliable-broadcast algorithms based on hashing and erasure coding~\cite{DBLP:journals/corr/abs-2007-14990}. These works are close in spirit to the present study because they treat reliable broadcast as executable software rather than only as pseudocode.

Artifact availability also matters for MBRB. An earlier AFRT implementation is available in Rust~\cite{AFRTRustImplementation}, and an earlier Coded MBRB prototype was implemented in Python with Merkle-tree-based authentication structures rather than vector commitments~\cite{DisatnikBoshoerCodedMBRB}. These artifacts are useful engineering evidence, but they do not by themselves provide a common comparison framework for Bracha, AFRT, and Coded MBRB under shared orchestration, workloads, metrics, and fault-injection semantics.

Our artifact is intended to narrow this implementation gap. Its contribution is not merely that three algorithms are implemented, but that they are implemented in one codebase with a common controller, persistent peer connections, manager-layer instrumentation, metric collection, parser checks, and fault-injection paths. This common infrastructure reduces confounding differences when comparing byte overhead, implementation-message count, CPU instructions, memory usage, latency, delivery behavior, and parser-checked safety outcomes.

\subsection{Reliable Dissemination inside Larger BFT Systems}
\label{sec:rw-bft-systems}

Reliable dissemination is also a core component of larger BFT systems. HoneyBadgerBFT uses reliable broadcast and asynchronous common subset machinery to build practical asynchronous atomic broadcast~\cite{DBLP:conf/ccs/MillerXCSS16}. Narwhal and Tusk separate reliable transaction dissemination from ordering, making dissemination a first-class bottleneck in high-throughput BFT system design~\cite{DBLP:conf/eurosys/DanezisKSS22}. Related DAG-based and asynchronous BFT systems similarly show that the cost of moving data reliably can dominate end-to-end system behavior even when consensus or ordering is the nominal abstraction.

These systems motivate the operational importance of reliable broadcast, but they are not direct experimental baselines for this paper. They evaluate complete BFT stacks with additional batching, mempool, ordering, consensus, cryptographic, and deployment machinery. Our target is narrower: standalone BRB and MBRB implementations evaluated under common single-shot workloads so that the cost of the reliable-broadcast primitive itself is visible.

\subsection{Benchmarking, Simulation, and Byzantine Testing}
\label{sec:rw-benchmarking-testing}

Systems work on BFT benchmarking and simulation provides important methodological context. BFT Protocols Under Fire showed that controlled experimentation can expose behavior not apparent from algorithm descriptions alone~\cite{DBLP:conf/nsdi/SinghDMDR08}. Later work on scalable BFT performance evaluation and simulation of unmodified BFT implementations demonstrates how network simulation can support reproducibility and larger-scale comparisons~\cite{DBLP:conf/prdc/BergerTR23,DBLP:journals/fac/BergerTR24}. This methodological line motivates our use of controlled simulation for network overhead and fault injection, complemented by native profiling and deployment-oriented latency measurements.

Byzantine testing tools provide a second methodological reference point. Twins generates Byzantine behaviors by duplicating identities and controlling communication partitions~\cite{DBLP:conf/opodis/BanoSCPLCM21}. ByzzBench benchmarks testing algorithms for BFT implementations~\cite{DBLP:conf/fmbc/NetoO25}. ByzzFuzz applies randomized testing and message perturbation to BFT implementations~\cite{DBLP:journals/pacmpl/WinterBGGO23}. BFTDiagnosis studies diagnosis and security indicators for BFT systems~\cite{DBLP:journals/cn/WangZWWH24}. These works support the importance of controlled adversarial testing, but their goals differ from benchmarking standalone reliable-broadcast implementations.

Our fault injection is deliberately narrower than these general Byzantine-testing frameworks. Silent nodes, randomized message omissions, and a partition-based equivocating sender are controlled experimental scenarios used to evaluate implementation behavior against expected thresholds and parser-checked safety conditions. They do not constitute an exhaustive Byzantine campaign, and they are not presented as a replacement for general-purpose BFT testing tools.

\subsection{Positioning of This Work}
\label{sec:rw-positioning}

The present paper sits between formal reliable-broadcast algorithms and full-system BFT benchmarking. It does not propose a new broadcast algorithm, a production-ready dissemination layer, or a complete benchmark standard. Instead, it asks how three existing algorithms behave when implemented, orchestrated, instrumented, and executed under comparable workloads and selected fault-injection scenarios. This perspective exposes costs that are abstracted away in pseudocode and asymptotic bounds: serialization, cryptographic verification, erasure coding, buffering, payload copying, memory allocation, and the semantics of injected faults.

This positioning also explains the algorithm selection. Bracha represents the classical asynchronous BRB baseline. AFRT represents signature-based message-adversary-tolerant broadcast. Coded MBRB represents a coded message-adversary-tolerant design that targets lower communication cost. Newer communication-efficient BRB algorithms, partially connected reliable-broadcast algorithms, large BFT stacks, and general Byzantine-testing frameworks are related but not direct experimental baselines. The artifact is intended to make such future comparisons easier by preserving a shared transport, manager, configuration, metric, parser, and plotting interface.

\section{Methodology}
\label{sec:detailed-methodology}

We define the evaluation methodology. Bracha {serves as a cross-abstraction BRB reference, whereas the AFRT--Coded MBRB comparison isolates implementation differences within MBRB. The evaluation is a systems study rather than a correctness proof. It therefore focuses on controlled comparability, reproducibility, resource costs, deployment-oriented latency, and behavior under selected injected faults. The implementations support single-shot broadcasts, isolating one invocation of the reliable-broadcast primitive from sequence-number management, batching, sliding windows, concurrent-instance scheduling, and multi-shot} garbage collection.

\subsection{Research Questions and Experiment Mapping}
\label{sec:rq-experiment-mapping}

The evaluation is organized around four research questions.

\paragraph*{RQ1: Communication scalability.}
How do the three implementations scale in transmitted data as the number of nodes and payload size increase, and how does the implementation-message count scale with the number of nodes? 

\paragraph*{RQ2: Local resource cost.}
How do CPU instruction count and memory usage scale, and when does Coded MBRB's cryptographic and coding cost become preferable to the full-payload processing costs of Bracha and AFRT?

\paragraph*{RQ3: Deployment latency.}
Do the communication savings of Coded MBRB translate into lower broadcast completion latency in deployment-oriented environments?

\paragraph*{RQ4: Fault-injection behavior.}
Do the implementations behave consistently with their expected safety, liveness, and threshold behavior under the tested silent-node, randomized message-omission, and partition-based equivocation scenarios?

\medskip
Table~\ref{tab:app-rq-experiment-mapping} maps each research question to the corresponding experiment, testbed, and metric group. The purpose of this table is to provide a detailed record supporting the three experiments reported in the paper.

\begin{table}[t]
\centering
\caption{Research-question to experiment mapping.}
\label{tab:app-rq-experiment-mapping}
\scriptsize
\setlength{\tabcolsep}{4pt}
\renewcommand{\arraystretch}{1.2}
\begin{tabularx}{\textwidth}{
  >{\hsize=0.3\hsize\raggedright\arraybackslash}X
  >{\hsize=1.2\hsize\raggedright\arraybackslash}X
  >{\hsize=0.9\hsize\raggedright\arraybackslash}X
  >{\hsize=1.8\hsize\raggedright\arraybackslash}X
  >{\hsize=0.8\hsize\raggedright\arraybackslash}X
}
\toprule
\textbf{RQ} & \textbf{Experiment} & \textbf{Testbed(s)} & \textbf{Measured quantities} & \textbf{Detailed evidence} \\
\midrule
RQ1 & Experiment~1: Controlled Scalability and Resource Costs & Shadow simulation & Total transmitted data node and payload scaling, and implementation-message count under node scaling & Section~\ref{sec:experiment1-results} \\
RQ2 & Experiment~1: Controlled Scalability and Resource Costs & Native profiling & User-space CPU instructions, peak heap, and cumulative allocation under node and payload scaling & Section~\ref{sec:experiment1-results} \\
RQ3 & Experiment~2: Deployment Latency & GCP and FABRIC & Controller-measured broadcast completion latency under payload and node scaling; RTT context for deployment measurements & Section~\ref{sec:experiment2-results} \\
RQ4 & Experiment~3: Fault Injection, Threshold Behavior, and Parser Checks & Shadow simulation, native profiling and parser pipeline & Delivery counts, delivery ratio, delivered hashes, sender hashes, duplicate-delivery flags, conflicting-delivery flags, spurious-delivery flags and validity flags & Sections~\ref{sec:experiment3-results} and~\ref{sec:parser-check-campaign} \\
\bottomrule
\end{tabularx}
\end{table}

\subsection{Orchestration and Node Architecture}
\label{sec:orchestration-node-architecture}

The benchmark consists of a centralized controller and a set of peer-to-peer nodes. The controller is not part of the reliable-broadcast algorithm being evaluated. It is used to distribute configuration, synchronize the beginning and end of each run, and collect metrics after the algorithm has terminated. This design makes the experiments repeatable and makes it possible to execute the same algorithms across Shadow, native profiling, GCP, and FABRIC.

\paragraph*{Controller lifecycle.}
At startup, the controller exposes an HTTP REST interface. Each node registers with the controller by sending its node identifier, public key, and listening address. The controller holds these registration requests until the expected number of nodes have registered. This registration barrier prevents a node from starting the experiment before the full network membership and cryptographic material are known.

Once all nodes have registered, the controller returns the configuration for each node. The configuration includes the selected algorithm, the peer list, payload size, cryptographic material, and fault-injection role. The controller validates configuration consistency, including that the requested injected faults do not exceed the run's configured \(t\) and \(d\) threshold values. After receiving the configuration, nodes establish pairwise peer-to-peer TCP connections. To avoid duplicate connections, a node initiates a connection only to peers with larger identifiers; this yields one bidirectional connection between each pair.

After the peer-to-peer network has been established, nodes notify the controller through a readiness endpoint. The controller again acts as a barrier and releases all nodes only when the expected number of nodes are ready. The designated sender then initiates the broadcast. Upon local delivery, each node reports completion to the controller. In fault-injection experiments, silent nodes are configured to signal completion even though they suppress algorithmic messages. After all expected completion signals have been received, the controller waits for a configurable grace period, set to three seconds by default, before asking nodes to upload metrics. The grace period allows in-flight algorithmic messages to be logged before the final metric snapshot is taken.

\begin{figure}
\centering
	\begin{tikzpicture}[
			scale=0.75, transform shape,
			>=Stealth,
			actor/.style={draw=black, thick, fill=white, minimum width=2.8cm, minimum height=0.8cm, font=\sffamily\bfseries},
			lifeline/.style={thick, dashed, draw=black},
			msg/.style={->, thick, draw=black},
			msg label/.style={above, font=\sffamily\scriptsize, align=center},
			action box/.style={draw=black, thick, fill=white, font=\sffamily\scriptsize, align=center, inner sep=6pt},
			black box/.style={draw=black, thick, fill=white, font=\sffamily\bfseries, align=center},
			barrier box/.style={draw=black, thick, fill=gray!10, font=\sffamily\scriptsize\bfseries, align=center}
		]

		\def\xC{0}
		\def\xNi{5.5}
		\def\xNj{11}
		\def\yEnd{-23}

		\node[actor] (C) at (\xC, 0) {Controller};
		\node[actor] (Ni) at (\xNi, 0) {Node $i$};
		\node[actor] (Nj) at (\xNj, 0) {Node $j$ ($j > i$)};

		\draw[lifeline] (C) -- (\xC, \yEnd);
		\draw[lifeline] (Ni) -- (\xNi, \yEnd);
		\draw[lifeline] (Nj) -- (\xNj, \yEnd);

		\node[action box] at (\xC, -1.2) {Start and listen for\\ requests};
		\node[action box] at (\xNi, -1.2) {Start P2P server\\(Prepare communication)};
		\node[action box] at (\xNj, -1.2) {Start P2P server\\(Prepare communication)};

		\draw[msg] (\xNi, -2.8) -- (\xC, -2.8) node[midway, msg label] {Register (ID, IP, Port, Public Key)};
		\draw[msg] (\xNj, -3.5) -- (\xC, -3.5) node[midway, msg label] {Register (ID, IP, Port, Public Key)};

		\filldraw[barrier box] (\xC-1.5, -4.3) rectangle (\xNj+1.5, -5.3)
		node[midway] {BARRIER: Wait for $n$ nodes to be registered};

		\draw[msg] (\xC, -6.5) -- (\xNi, -6.5) node[midway, msg label] {Config (Algorithm, Role, Peer List, Crypto, ...)};
		\draw[msg] (\xC, -7.2) -- (\xNj, -7.2) node[midway, msg label] {Config (Algorithm, Role, Peer List, Crypto, ...)};

		\draw[msg] (\xNi, -8.5) -- (\xNj, -8.5) node[midway, msg label] {Establish P2P Connection ($i < j$)};

		\draw[msg] (\xNi, -9.8) -- (\xC, -9.8) node[midway, msg label] {Ready};
		\draw[msg] (\xNj, -10.5) -- (\xC, -10.5) node[midway, msg label] {Ready};

		\filldraw[barrier box] (\xC-1.5, -11.5) rectangle (\xNj+1.5, -12.5)
		node[midway] {BARRIER: Wait for $n$ nodes to be ready};

		\filldraw[black box] (\xNi-1.5, -13.5) rectangle (\xNj+1.5, -15.5)
		node[midway] {Algorithm execution};

		\node[action box] at (\xNi, -16.5) {Local Delivery};
		\node[action box] at (\xNj, -16.5) {Local Delivery};

		\draw[msg] (\xNi, -17.8) -- (\xC, -17.8) node[midway, msg label] {Done};
		\draw[msg] (\xNj, -18.5) -- (\xC, -18.5) node[midway, msg label] {Done};

		\filldraw[barrier box] (\xC-1.5, -19.5) rectangle (\xNj+1.5, -20.5)
		node[midway] {BARRIER: Wait for $n$ nodes to be done};

		\draw[msg] (\xNi, -21.5) -- (\xC, -21.5) node[midway, msg label] {Push collected metrics};
		\draw[msg] (\xNj, -22.2) -- (\xC, -22.2) node[midway, msg label] {Push collected metrics};

	\end{tikzpicture}
\caption{Detailed methodology: orchestration sequence used by the benchmark. The controller acts as a registration, readiness, completion, and metric-collection barrier, while algorithmic messages are exchanged directly among nodes.}
\label{fig:app-systemcommunication}
\end{figure}
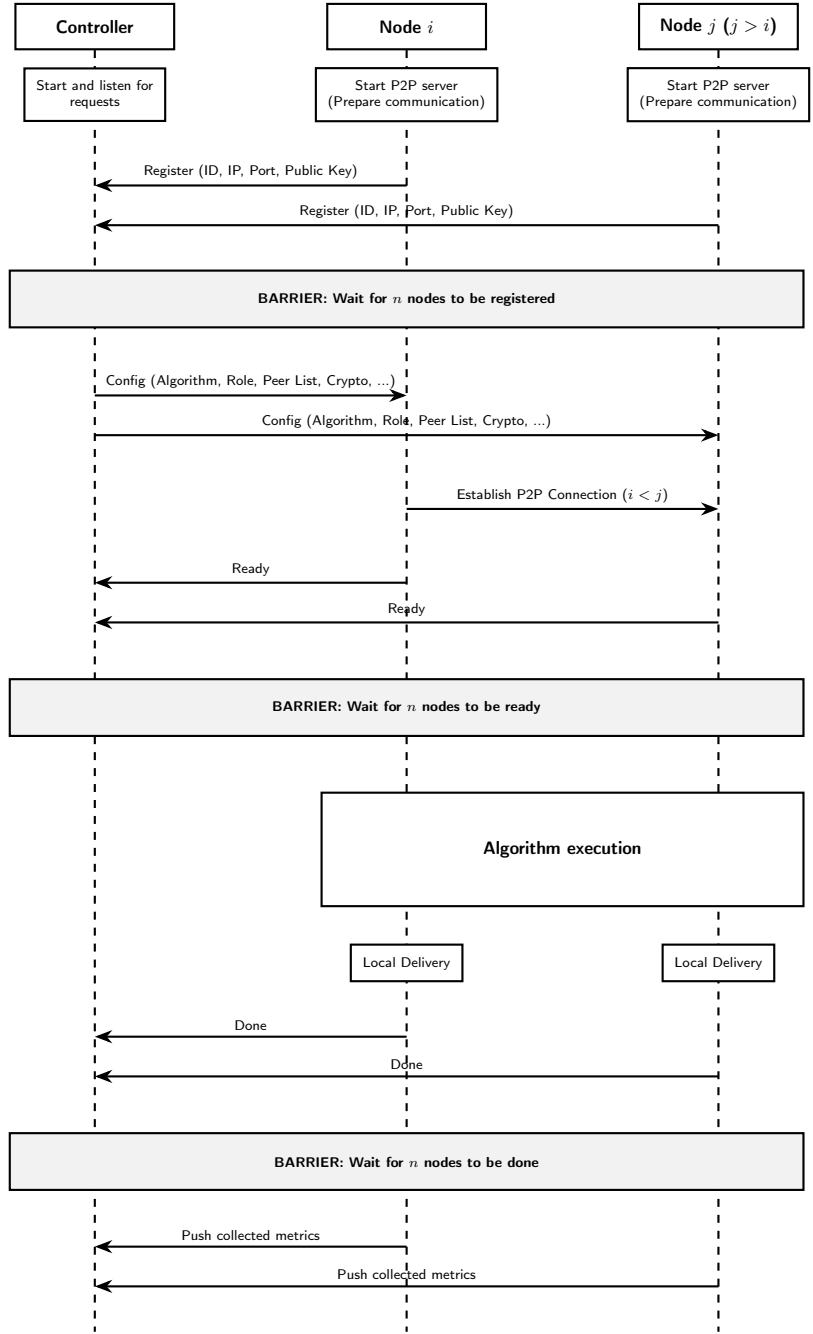

\paragraph*{Node layers.}
Each node separates transport, management, metrics, and algorithm logic. The Transport layer establishes the peer-to-peer connections and provides byte-stream send and receive functionality. The Manager layer exposes two primitives to the algorithm implementation: a broadcast primitive for sending the same implementation message to all peers, and a per-recipient send primitive for sending distinct implementation messages to selected peers. The algorithm layer is responsible for serializing and deserializing algorithm-specific message structures.

The Manager layer is also the common measurement and interception point. It records sent and received implementation messages, byte counts, and delivery timestamps. It also hosts the fault-injection wrappers used to suppress outgoing or incoming messages in selected experiments. This design is important for comparability: Bracha, AFRT, and Coded MBRB are evaluated through the same orchestration, transport, metric, and fault-injection interfaces.

\paragraph*{Metric snapshot.}
At the end of each execution, every node reports a structured metric object to the controller. This object contains the node identifier, role, delivery status, delivery count, sent and received byte counts, sent and received implementation-message counts, algorithm initiation and delivery timestamps, memory statistics, and hashes. If the node delivered a payload, it reports the SHA-256 hash of the delivered payload. The designated sender additionally reports the SHA-256 hash of the original application payload. These hashes are used by the parser to detect inconsistent or invalid delivery outcomes.

\subsection{Implementation Details Affecting Evaluation}
\label{sec:implementation-evaluation-details}

This subsection records implementation choices that directly affect measurement validity or interpretation. It is not intended as a complete implementation manual.

\paragraph*{Concurrency and locking.}
The algorithm implementations originally processed messages sequentially. To reduce avoidable latency, the implementation was later modified to process incoming messages concurrently. Algorithm state, such as collected signatures, fragments, quorums, and delivery status, is kept in shared state protected by mutual exclusion locks. Critical sections were kept as short as possible, especially for AFRT and Coded MBRB, whose cryptographic operations can be expensive. When possible, a node copies the state needed for a computation while holding the lock, releases the lock, and then performs the heavier cryptographic or coding operation outside the critical section.

\paragraph*{Serialization.}
The algorithms serialize their own implementation messages before passing byte streams to the Manager and Transport layers. Early versions used JSON serialization. The final evaluation uses Go's \texttt{gob} encoding because JSON introduced substantial encoding overhead and additional processing cost. This matters for the network-overhead metric: the measured bytes are serialized implementation messages, including algorithmic metadata, signatures, fragments, commitments, and payload data when applicable.

\paragraph*{Cryptographic choices.}
AFRT uses Go's \texttt{crypto/ed25519} package for public-key signatures. Ed25519 was selected because it is simple to use and provides compact signatures. Coded MBRB uses \texttt{gnark-crypto} for KZG vector commitments and threshold-signature operations over BN254, and \texttt{klauspost/reedsolomon} for erasure coding. These choices affect both CPU instruction count and memory behavior. In particular, the high fixed CPU cost observed for Coded MBRB is tied to generating vector commitments, verifying inclusion proofs, producing and combining threshold-signature shares, and encoding or reconstructing fragments.

\paragraph*{Measurement boundary.}
The byte metric is collected immediately before an implementation message is passed from the Manager layer to the Transport layer. It therefore includes serialized algorithmic content but excludes TCP/IP headers, HTTP orchestration messages, and metric-upload traffic. CPU measurements in native profiling cover the node process from initialization to termination; because the same orchestration path is used for all algorithms, the common initialization and termination overheads are present across all measurements. Memory measurements are collected through Go's runtime after the controller releases the termination barrier.

\subsection{Single-Shot Scope}
\label{sec:single-shot-scope}

The implementation evaluates single-shot broadcast instances. Each run consists of one invocation of the selected reliable-broadcast algorithm for one application payload. This design intentionally isolates the implementation cost of the reliable-broadcast primitive itself. A multi-shot system would require additional mechanisms such as sequence numbers, batching, sliding windows, admission control, and garbage collection across many concurrent or repeated broadcast instances.

The single-shot design is therefore a methodological choice rather than a claim that deployed systems use reliable broadcast only once. It avoids confounding the measurements with wrapper-level engineering decisions. For example, a stop-and-wait wrapper would limit throughput to roughly one completed instance per round-trip time, while a sliding-window design would introduce buffer-management policies and possible resource-exhaustion behavior. Those effects are important for future work, but they are not part of the per-broadcast costs measured here.

\subsection{Evaluation Environments}
\label{sec:evaluation-environments}

Table~\ref{tab:app-environment-methods} summarizes the detailed role of each evaluation environment. The environments are complementary. Shadow provides controlled and reproducible network and fault-injection behavior. Native profiling isolates CPU and memory costs that Shadow does not accurately capture. GCP provides the primary deployment-oriented latency evidence. FABRIC provides supplementary latency evidence over a more distributed infrastructure.
Unless otherwise stated, each configuration is repeated 10 times.

\begin{table}[ht]
\centering
\caption{Evaluation environments and their role in the benchmark. Because the benchmark uses a shared manager layer, all executions generate a unified JSON schema containing: algorithm, $n$, $t$, $d$, payload size, fault configurations, controller timestamps, delivery status, hashes, message/byte counts, and memory statistics.}
\label{tab:app-environment-methods}
\scriptsize
\setlength{\tabcolsep}{3pt}
\renewcommand{\arraystretch}{1.2}
\begin{tabularx}{\textwidth}{
> {\hsize=0.65\hsize\raggedright\arraybackslash}X
> {\hsize=1.15\hsize\raggedright\arraybackslash}X
> {\hsize=0.85\hsize\raggedright\arraybackslash}X
> {\hsize=0.9\hsize\raggedright\arraybackslash}X
> {\hsize=1.15\hsize\raggedright\arraybackslash}X
> {\hsize=1.15\hsize\raggedright\arraybackslash}X
}
\toprule
\textbf{Env.} & \textbf{Setup} & \textbf{Parameters varied} & \textbf{Measured quantities} & \textbf{Purpose} & \textbf{Main limitation} \\
\midrule
Shadow simulation
& Shadow v3.3.0 in Ubuntu 24.04 container, simulated 1~Gbps switch
& $n$, payload size, silent nodes, message drops, sender equivocation
& Transmitted bytes, implementation messages, delivery counts
& Controlled communication and fault-injection experiments using the same Go binaries
& Simulated time does not capture local CPU delay \\
Native profiling
& Fedora 44 x86-64 localhost, Linux \texttt{perf}, Go \texttt{ReadMemStats}
& $n$, payload size, silent nodes
& User-space CPU instructions, peak heap, cumulative allocation
& Isolate local processing and memory costs outside Shadow
& Localhost execution does not model network latency \\
Google Cloud Platform
& VM deployment with private VPC communication
& $n$, payload size
& Controller-measured broadcast completion latency
& Deployment-oriented latency under Virtual Machine (VM) and Virtual Private Cloud (VPC) execution
& Small cloud deployment; not a production or wide-area study \\
FABRIC testbed
& Distributed multi-site testbed deployment
& $n$, payload size
& Controller-measured broadcast completion latency
& Supplementary deployment-oriented latency evidence over a distributed infrastructure
& Site placement and RTT distribution affect comparability \\
\midrule
\textit{Post-processing consistency checks}
& --
& \textit{All generated runs}
& \textit{Delivery count, delivered hash, sender hash, violation flags}
& \textit{Checking of safety-related outcomes and aggregation of raw JSON outputs}
& \textit{Checks executions generated by the test campaign; not exhaustive verification} \\
\bottomrule
\end{tabularx}
\end{table}

\paragraph*{Shadow simulation.}
The Shadow environment executes the compiled Go binaries over a simulated network topology. All nodes are connected through a simulated 1~Gbps switch. For each Shadow configuration, the experiment is repeated with 10 recorded seeds: \{42, 286, 386, 407, 486, 1337, 6502, 8086, 25565, 68000\}. This matters because the message-omission fault injector samples recipients randomly.

\paragraph*{Native profiling.}
The native profiling environment runs the same binaries on localhost. CPU instructions are collected with \texttt{perf stat -e instructions:u} and appended to the metric JSON. The \texttt{:u} suffix restricts counting to user-space instructions. Memory statistics are collected using Go's \texttt{runtime.ReadMemStats}. The main memory metrics are peak heap, reported through \texttt{HeapSys}, and cumulative allocation, reported through \texttt{TotalAlloc}.

\paragraph*{Google Cloud Platform.}
The GCP deployment is the primary deployment-oriented latency environment. Nodes communicate over a private VPC. The controller measures completion latency as the elapsed time between releasing the ready barrier and receiving completion signals from all honest nodes. This avoids relying on synchronized clocks across VMs. In the GCP deployment, RTT measurements among nodes produced average RTTs between 0.25~ms and 0.86~ms, with peak delays below 3~ms.

\paragraph*{FABRIC.}
FABRIC is retained as supplementary deployment evidence. It is useful because it exercises a more distributed infrastructure and supports larger latency experiments than the small GCP setup. However, FABRIC measurements must be interpreted together with the placement of nodes across the five FABRIC sites used (TACC, UTAH, NCSA, MAX, and MICH) and the measured RTT distribution. The RTT between the nodes was measured before every single broadcast execution during the initial P2P connection phase. These measurements were conducted across the approximately 3 hours and 50 minutes it took to execute all performed algorithm tests on FABRIC. The results of these measurements are detailed in Section~\ref{sec:fabric-results}.

\subsection{Data Collection Pipeline}
\label{sec:data-collection-pipeline}

The evaluation workflow is automated by scripts included in the public artifact. The controller aggregates node-level metrics into one JSON file per run. Each JSON file records both node metrics and global experiment parameters, including algorithm, network size, Byzantine threshold $t$, message-adversary power $d$, payload size, sender identifier, number of silent nodes, number of message drops, and controller-measured start and end timestamps.

Environment-specific metadata is appended to the raw output by these scripts. Shadow outputs include the seed used for the run, while native profiling and deployment-oriented outputs include the iteration index. Native-profiling includes parsed \texttt{perf} instruction counts for each node.

A secondary parser converts raw JSON files into structured CSV files and checks safety-related outcomes. The parser checks whether an honest node delivered more than once, whether delivering honest nodes delivered different hashes, whether the delivered hash matches the sender hash when the sender is correct, and whether a node delivered when no broadcast was initiated. These checks are deterministic post-processing checks over the generated experiment outputs; they are not formal verification and they do not exhaustively explore all Byzantine executions.

Plots are generated after two aggregation stages. First, node metrics are aggregated within each broadcast execution. For example, total transmitted data is summed over nodes, while CPU instruction count can be averaged per node. Second, the resulting execution-level values are aggregated over the 10 repetitions for the configuration. Unless explicitly stated otherwise, figures report the arithmetic mean over 10 runs, and shaded regions represent one standard deviation. We do not remove outliers because the observed distributions were consistently narrow.

\subsection{Metric Definitions}
\label{sec:metric-definitions}

This subsection gives the metric boundaries used throughout the paper.

\paragraph*{Network overhead.}
Let $B_i$ denote the number of bytes of implementation messages sent by node $i$ during one broadcast instance. Total network overhead is $\sum_i B_i$. The metric includes serialized implementation messages encoded with \texttt{gob}; this includes payloads, fragments, signatures, vector commitments, inclusion proofs, threshold-signature shares, and other algorithmic metadata. It excludes TCP/IP headers, HTTP orchestration traffic, and metric-upload traffic.

\paragraph*{Implementation-message count.}
Implementation-message count is the number of algorithm-generated messages sent by the nodes. This metric is intentionally separate from total transmitted bytes. Coded MBRB may transmit many lightweight implementation messages while still transmitting substantially fewer bytes than full-payload algorithms.

\paragraph*{CPU instruction count.}
Computational cost is measured as user-space CPU instructions per node using \texttt{perf stat -e instructions:u}. Because this profiles the full node process from initialization to termination, the measurement includes common orchestration overhead. However, the same initialization and termination path is used across algorithms, making the metric useful for relative comparison.

\paragraph*{Memory usage.}
Peak heap is reported using Go's \texttt{HeapSys}. Cumulative allocation is reported using \texttt{TotalAlloc}. Peak heap captures the maximum memory pressure reached during the execution, while cumulative allocation captures allocation churn during the run. We emphasize peak heap, but we include cumulative allocation because it helps explain payload-copying and buffering costs.

\paragraph*{Broadcast completion latency.}
Deployment latency is measured at the controller. Let $t_{start}$ be the time at which the controller releases the ready barrier, and let $t_{end}$ be the time at which the controller has received completion signals from all honest nodes. Completion latency is $t_{end}-t_{start}$. This avoids clock-skew artifacts but includes controller-node transit time.

\paragraph*{Delivery ratio.}
Let $H$ be the set of honest nodes and let $D \subseteq H$ be the honest nodes that deliver before the timeout. The delivery ratio is $|D|/|H|$. In the fault-threshold experiments, the absolute numbers of delivering nodes are often reported because the theoretical guarantees for MBRB are stated in terms of delivery power.

\paragraph*{Safety checks.}
The parser records delivery count, delivered hash, and sender hash. A duplicate-delivery violation is flagged if an honest node delivers more than once. A conflicting-delivery violation is flagged if two honest nodes deliver different hashes for the same broadcast. A validity violation is flagged if a delivered hash differs from the sender hash in a correct-sender execution. A spurious-delivery violation is flagged if a node delivers when no broadcast was initiated.

\subsection{Fault-Injection Mechanisms}
\label{sec:fault-injection}

Fault injection is implemented at the Manager layer so that the honest algorithm code remains unchanged.

\paragraph*{Silent nodes.}
A silent node remains alive but suppresses incoming and outgoing algorithmic messages. This differs from killing the process because killing the process would cause the operating system to close TCP connections and potentially reveal the failure at the transport layer. The silent-node mechanism instead approximates an undetectable crash from the algorithm's perspective.
This number of silent nodes is configured by the \texttt{numSilent} parameter.

\paragraph*{Randomized message omission.}
To simulate message omission, the Manager layer uses a parameter \texttt{numMsgDrops}. For each implementation-message broadcast, it uniformly samples up to \texttt{numMsgDrops} recipients among honest nodes without replacement and suppresses those outgoing messages before they reach the transport layer. This models message omission at the implementation-message level rather than packet loss at the TCP level. It is not a worst-case adaptive message adversary because it does not select messages based on global real-time knowledge of the algorithm state.

\paragraph*{Partition-based equivocation.}
The equivocation experiment is Twins~\cite{DBLP:conf/opodis/BanoSCPLCM21} inspired. The sender runs two algorithm instances through two \texttt{SplitManager} wrappers, each restricted to one partition of the network. The two instances broadcast different payloads to different halves of the network. This creates a controlled split-sender equivocation scenario without modifying the honest algorithm code. The experiment does not exhaustively characterize Byzantine behavior; it tests one reproducible scenario designed to expose conflicting-delivery risks.
The partition-based equivocation is enabled by the Boolean parameter \texttt{twinsSender}.

\paragraph*{Configured thresholds and injected faults.}
The benchmark separates the configured mathematical fault thresholds from the actual injected faults. For Bracha, the relevant threshold condition is $n>3t$. For AFRT and Coded MBRB, the relevant threshold condition is $n>3t+2d$, where $d$ is the message-adversary removal power. The experiment configuration records both the tolerated parameters and the actually injected numbers of silent nodes and dropped messages.
Specifically, our threshold tests use $t = \texttt{numSilent}$ and $d = \texttt{numMsgDrops}$, and vary these values across configurations.

The injected faults target the two adversarial dimensions of our model: Byzantine-node behavior (silence and equivocation) and message-adversary omissions. In-transit modification of messages from correct nodes is outside the omission-only MA model, while AFRT and Coded MBRB additionally use cryptographic signatures to authenticate protocol evidence.

\subsection{Experimental Configurations}
\label{sec:experimental-configurations}

The evaluation comprises three RQ-mapped experiments. Table~\ref{tab:app-experiment-configurations} summarizes the configurations used by the three experiments. The results in Section~\ref{sec:results} follow the same experiment numbering.

\begin{table}[t]
\centering
\caption{Experiment configurations.}
\label{tab:app-experiment-configurations}
\scriptsize
\setlength{\tabcolsep}{4pt}
\renewcommand{\arraystretch}{1.2}
\begin{tabularx}{\textwidth}{
  >{\hsize=1.1\hsize\raggedright\arraybackslash}X
  >{\hsize=0.4\hsize\raggedright\arraybackslash}X
  >{\hsize=0.8\hsize\raggedright\arraybackslash}X
  >{\hsize=1.9\hsize\raggedright\arraybackslash}X
  >{\hsize=0.8\hsize\raggedright\arraybackslash}X
}
\toprule
\textbf{Experiment} & \textbf{RQs} & \textbf{Testbed(s)} & \textbf{Variables} & \textbf{Repetitions} \\
\midrule
Experiment~1: Controlled Scalability and Resource Costs & RQ1, RQ2 & Shadow and native profiling & Node scaling: $n=10,15,20,25,30$ at 1~MB; payload scaling: 100~KB--1~MB and 1--8~MB at $n=30$; extended CPU payload test: 10--40~MB at $n=10$ & 10 per configuration \\
Experiment~2: Deployment Latency & RQ3 & GCP and FABRIC & GCP payload scaling at $n=5$; GCP node scaling $n=4$--$11$ at 1~MB; FABRIC node and payload scaling as supplementary evidence & 10 per configuration \\
Experiment~3: Fault Injection, Threshold Behavior, and Parser Checks & RQ4 & Shadow, native profiling, and parser pipeline & Silent nodes, randomized message omissions, partition-based equivocation, and additional parser-check permutations & 10 per configuration \\
\bottomrule
\end{tabularx}
\end{table}

\noindent \textbf{Exp-1:} \emph{Controlled scalability and resource costs (RQ1 and RQ2).~~}
This experiment runs without fault injection and measures communication, CPU, and memory costs. In Shadow, we measure total transmitted data and implementation-message count. In native profiling, we measure CPU instructions and memory usage. We use two primary scaling dimensions: node scaling from $n=10$ to $30$ in increments of 5 with a fixed 1~MB payload, and payload scaling at $n=30$ from 0.1~MB to 8~MB. An additional native profiling experiment scales payloads from 10~MB to 40~MB at $n=10$ because 30-node configurations exceeded the host machine's memory limits at these larger payload sizes.

\noindent \textbf{Exp-2:} \emph{Deployment latency (RQ3).~~}
We measure broadcast completion latency in deployment-oriented environments. The GCP experiments evaluate payload scaling up to 8~MB at $n=5$ and node scaling from $n=4$ to $11$. Supplementary FABRIC experiments extend the deployment study to larger network sizes (up to $n=30$) and broader payload ranges (100~KB to 40~MB).

\noindent \textbf{Exp-3:} \emph{Fault injection and consistency checks (RQ4).~~}
This experiment runs in Shadow and native profiling at $n=30$ with a 1~MB payload for the main plotted delivery and parser-check results and tests varying numbers of silent nodes (\texttt{numSilent}), randomized message omissions (\texttt{numMsgDrops}), and the partition-based sender equivocation scenario (\texttt{twinsSender}).
{For the threshold tests, we set $t=\texttt{numSilent}$ and $d=\texttt{numMsgDrops}$. We compare observed delivery with the sufficient bounds $n>3t$ for Bracha and $n>3t+2d$ for AFRT and Coded MBRB.
The reported threshold scenarios use $(t,d)=(9,0),(10,0)$ for Bracha and $(5,7),(7,5)$ for AFRT and Coded MBRB; Section~\ref{sec:parser-check-campaign} details broader fault permutations and parser checks.}

Table~\ref{tab:app-parser-campaign} summarizes the parser-check accounting. The parser-check campaign is important because it connects the raw metric outputs to the safety-related claims made in the paper.

\begin{table}[ht]
\centering
\caption{Parser-check campaign accounting. The parser found zero duplicate-delivery, conflicting-delivery, invalid-delivery, or spurious-delivery violations in these checked outputs.}
\label{tab:app-parser-campaign}
\scriptsize
\setlength{\tabcolsep}{4pt}
\renewcommand{\arraystretch}{1.2}
\begin{tabularx}{\textwidth}{
  >{\hsize=1.2\hsize\raggedright\arraybackslash}X
  >{\hsize=0.4\hsize\raggedleft\arraybackslash}X
  >{\hsize=0.6\hsize\raggedleft\arraybackslash}X
  >{\hsize=1.8\hsize\raggedright\arraybackslash}X
}
\toprule
\textbf{Output group} & \textbf{Runs} & \textbf{Checked entries} & \textbf{Purpose} \\
\midrule
Shadow plots & 1,500 & 43,500 & Network overhead, message counts, and selected fault-injection plots \\
Native profiling plots & 840 & 19,500 & CPU and memory measurements \\
GCP deployment plots & 480 & 3,000 & GCP latency measurements \\
FABRIC deployment plots & 870 & 24,600 & FABRIC latency measurements \\
Native profiling fault-injection tests & 900 & 27,000 & CPU and memory degradation under silent nodes \\
Additional Shadow fault permutations & 87,600 & 2,244,000 & Additional silent-node, message-drop, and equivocation parser checks \\
\midrule
\textbf{Total} & \textbf{92,190} & \textbf{2,361,600} & Full parser-check campaign \\
\bottomrule
\end{tabularx}
\end{table}

\section{Evaluation Results}
\label{sec:results}

Across all generated outputs and test configurations, the parser checked 2,361,600 entries over 92,190 runs and found zero parser-detected property violations. These checks provide implementation-level evidence over the tested executions; they do not replace the formal correctness arguments of the algorithms and do not constitute exhaustive Byzantine testing.
{For the payload-dependent communication term, Bracha and AFRT incur $O(n^2|m|)$ total communication, whereas Coded MBRB targets $O(n|m|+n^2\kappa)$~\cite{DBLP:conf/opodis/AlbouyFGHRSTZ24,DBLP:journals/tcs/AlbouyFRT23,DBLP:journals/iandc/Bracha87}. Exp-1 examines whether these asymptotic differences appear in the implementations.}
The results expose a consistent system trade-off. Coded MBRB reduces transmitted data, peak heap, and deployment latency when payload movement or network transit dominates. This advantage comes at a higher CPU cost due to erasure coding, vector commitments, inclusion proofs, and threshold-signature operations. Bracha and AFRT are computationally lighter for smaller payloads and local workloads, but their repeated full-payload dissemination becomes expensive as the system or payload grow.

\subsection{Exp-1: Controlled Scalability and Resource Costs}
\label{sec:experiment1-results}

Experiment~1 answers RQ1 and RQ2. It uses Shadow simulation for communication cost and native profiling for CPU and memory cost. No faults are injected in this experiment.

\begin{figure}[ht]
\centering
\includegraphics[width=0.78\linewidth]{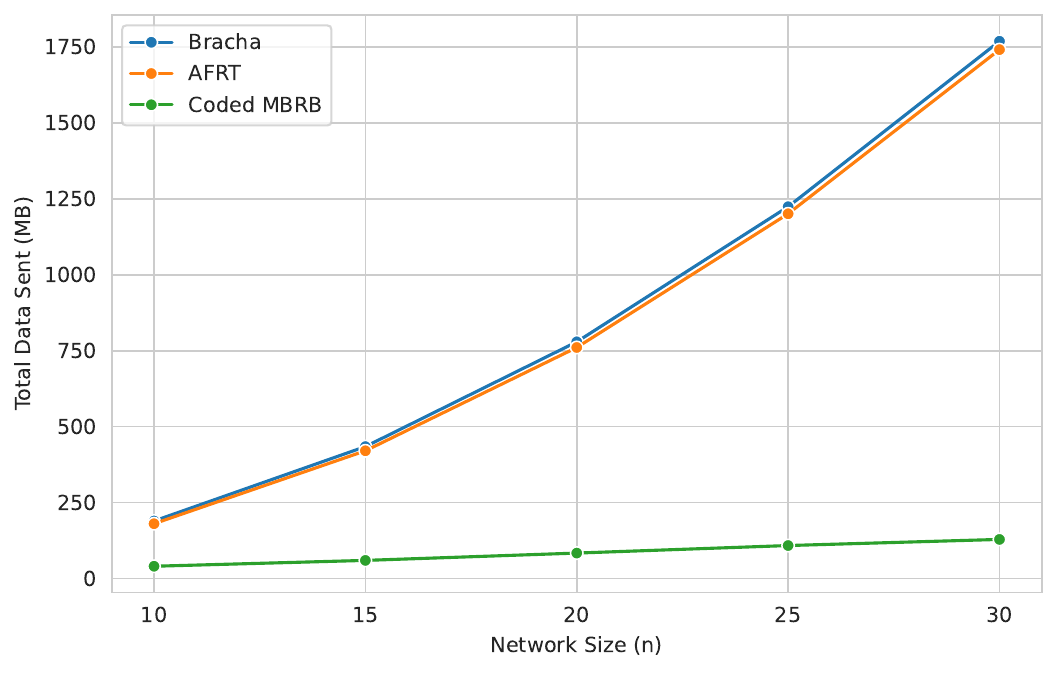}
\caption{Exp-1, Shadow: total transmitted data under node scaling with a 1~MB payload. Coded MBRB transmits substantially fewer bytes as $n$ grows.}
\label{fig:app-sim-data-sent-vs-nodes}
\end{figure}

\paragraph*{Shadow: total transmitted data under node scaling.}
Fig.~\ref{fig:app-sim-data-sent-vs-nodes} reports total transmitted data as the number of nodes increases from $n=10$ to $n=30$ with a fixed 1~MB payload. Bracha and AFRT increase much more aggressively than Coded MBRB. At $n=30$, Bracha and AFRT transmit roughly 1.7--1.8~GB per single broadcast, while Coded MBRB remains close to 128~MB. This is the central communication-scaling result for RQ1.

\begin{figure}[ht]
\centering
\includegraphics[width=0.78\linewidth]{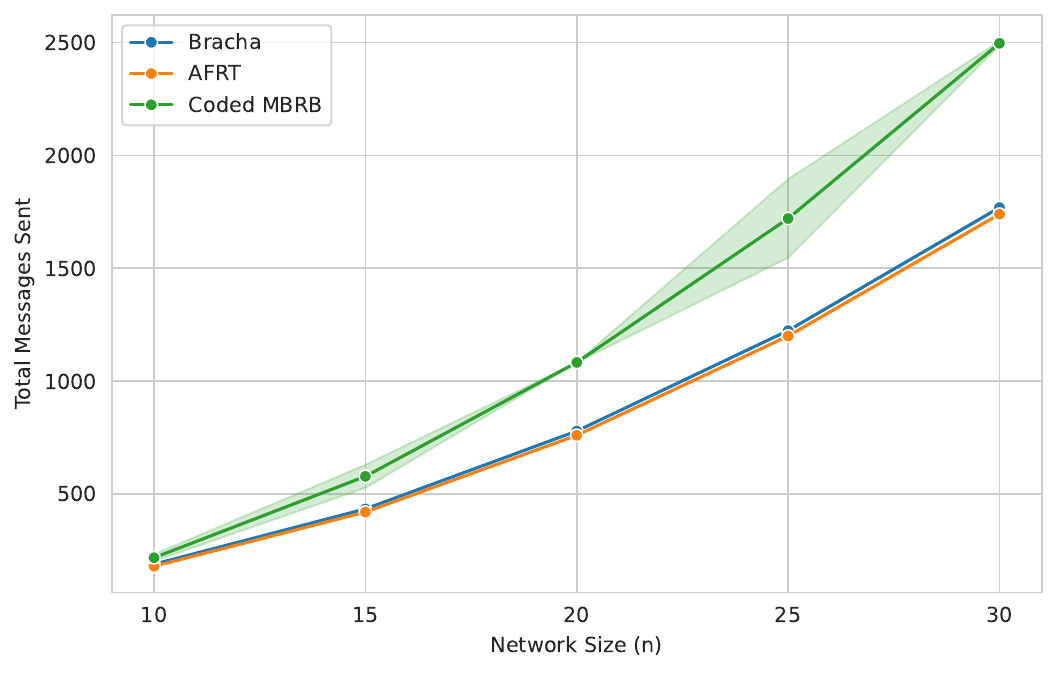}
\caption{Exp-1, Shadow: implementation-message count under node scaling with a 1~MB payload. Coded MBRB sends more lightweight implementation messages even while transmitting fewer bytes.}
\label{fig:app-sim-messages-vs-nodes}
\end{figure}

\paragraph*{Shadow: implementation-message count under node scaling.}
Fig.~\ref{fig:app-sim-messages-vs-nodes} reports the number of implementation messages under the same node-scaling configuration. Coded MBRB sends more individual implementation messages than Bracha and AFRT, reaching roughly 2500 messages at $n=30$, compared with roughly 1750 messages for Bracha and AFRT. This result is important because it separates byte scalability from message-count scalability. Coded MBRB improves total transmitted data, but not because it sends fewer messages; rather, it sends more lightweight messages containing fragments and cryptographic evidence.
Thus, Coded MBRB improves byte scalability, but not message-count.

\begin{figure}[ht]
\centering
\begin{subfigure}{0.48\textwidth}
\centering
\includegraphics[width=\textwidth]{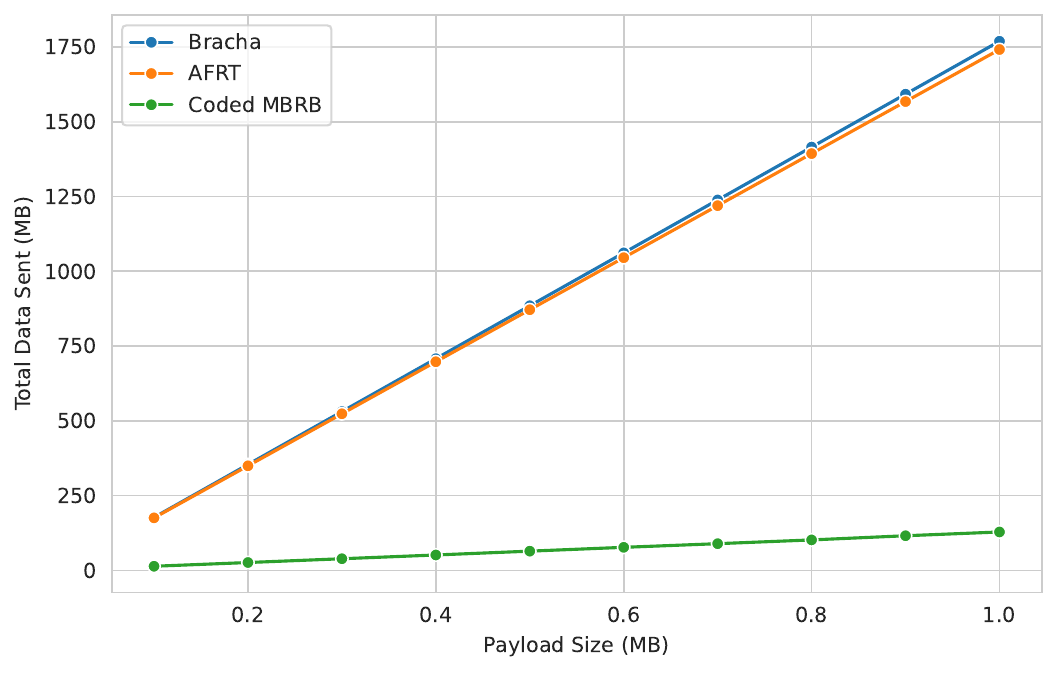}
\caption{100~KB--1~MB}
\end{subfigure}
\hfill
\begin{subfigure}{0.48\textwidth}
\centering
\includegraphics[width=\textwidth]{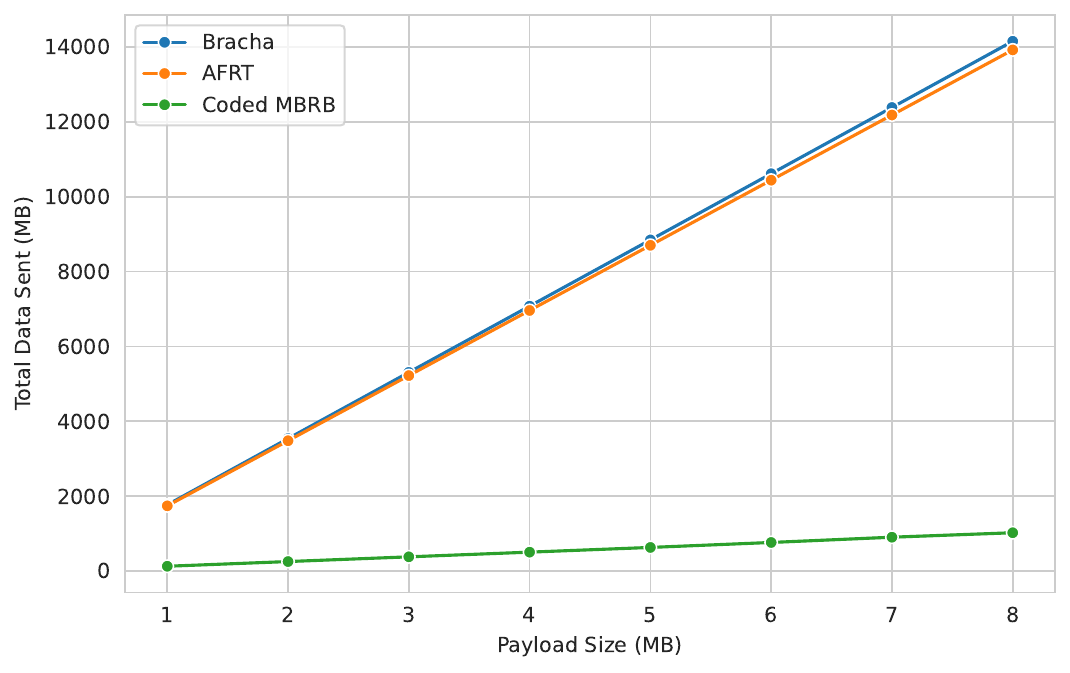}
\caption{1--8~MB}
\end{subfigure}
\caption{Exp-1, Shadow: total transmitted data under payload scaling at $n=30$. Full-payload dissemination dominates Bracha and AFRT as payload size grows.}
\label{fig:app-sim-data-sent-vs-payload}
\end{figure}

\paragraph*{Shadow: total transmitted data under payload scaling.}
Fig.~\ref{fig:app-sim-data-sent-vs-payload} reports total transmitted data at $n=30$ while payload size grows. Bracha and AFRT grow approximately linearly with payload size because they repeatedly disseminate the full payload. Coded MBRB grows much more slowly because it disseminates coded fragments and compact cryptographic evidence. At an 8~MB payload, Bracha and AFRT transmit approximately 14~GB per single broadcast, while Coded MBRB is near 1~GB.

\begin{figure}
\centering
\includegraphics[width=0.78\linewidth]{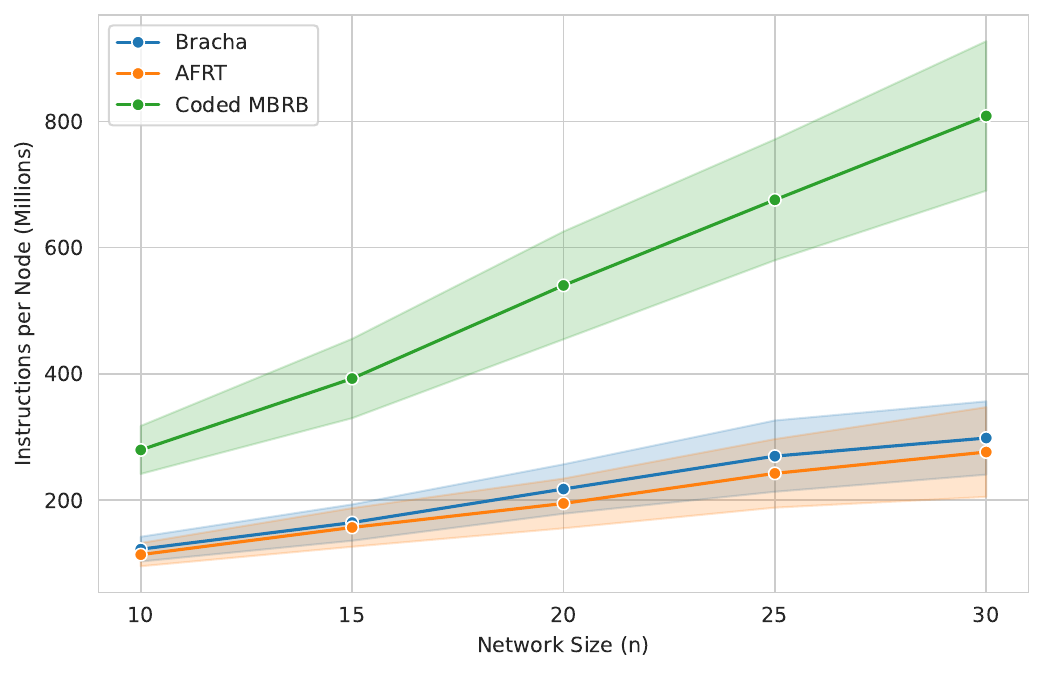}
\caption{Exp-1, native profiling: CPU instruction count under node scaling with a 1~MB payload. Coded MBRB pays a higher local computational cost due to coding and cryptographic operations.}
\label{fig:app-cpu-vs-nodes}
\end{figure}

\paragraph*{Native profiling: CPU instructions under node scaling.}
Fig.~\ref{fig:app-cpu-vs-nodes} reports user-space CPU instructions per node while the network size grows with a fixed 1~MB payload. At $n=30$, Coded MBRB executes approximately 800 million instructions per node, compared with roughly 250 million for Bracha and AFRT. This is the main counterweight to the communication result: Coded MBRB's byte savings are purchased with substantially higher local computation.

\begin{figure}
\centering
\begin{subfigure}{0.48\textwidth}
\centering
\includegraphics[width=\textwidth]{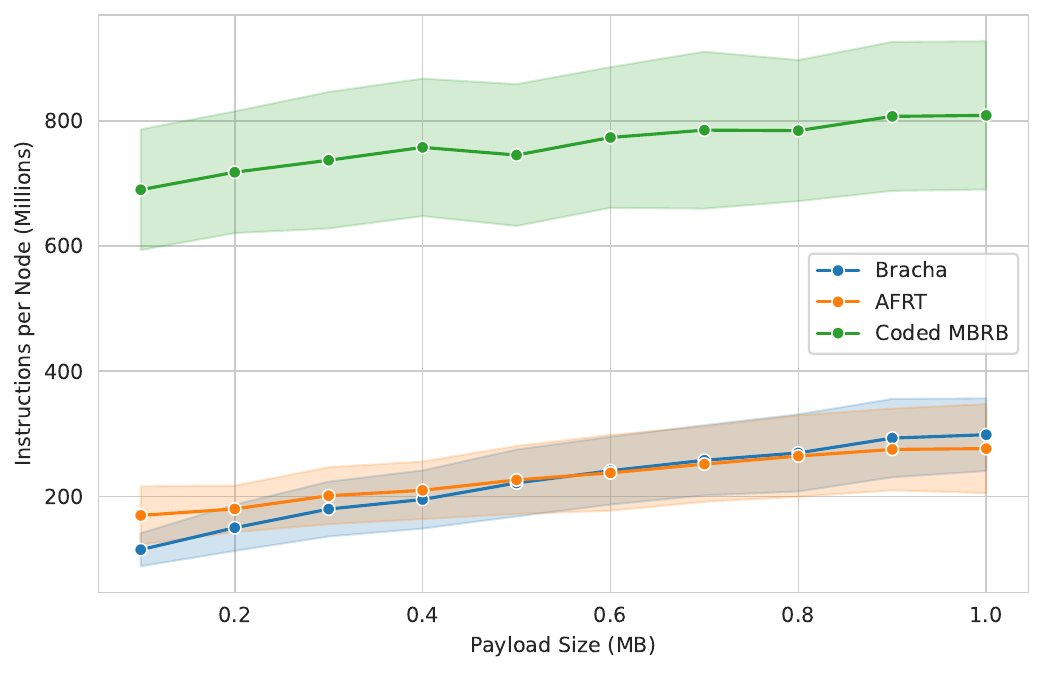}
\caption{100~KB--1~MB}
\end{subfigure}
\hfill
\begin{subfigure}{0.48\textwidth}
\centering
\includegraphics[width=\textwidth]{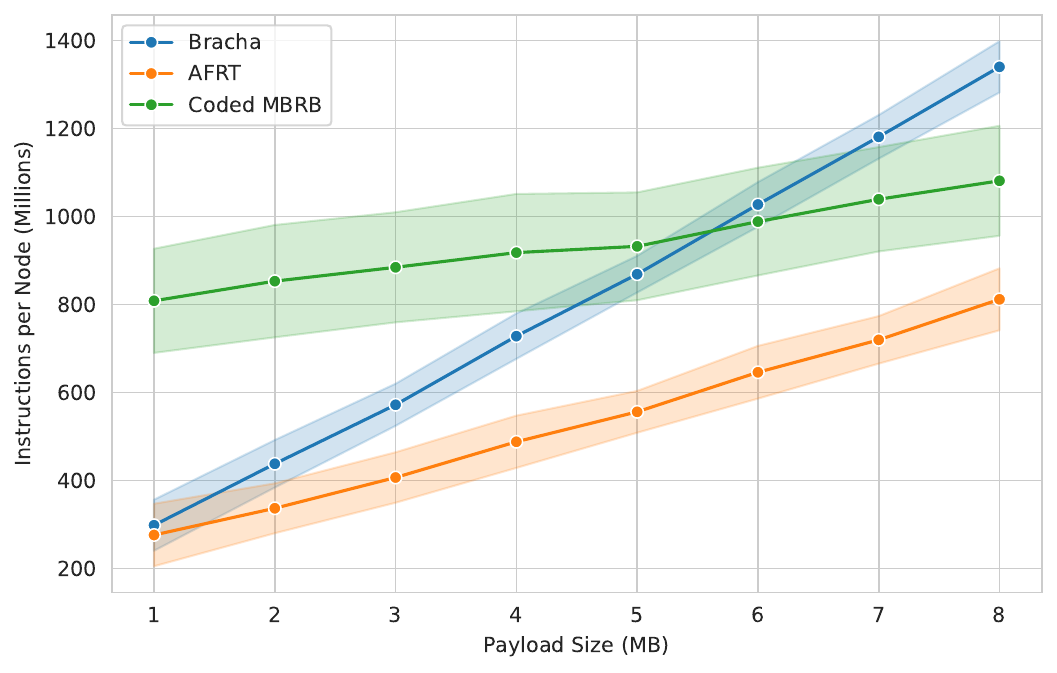}
\caption{1--8~MB}
\end{subfigure}
\caption{Exp-1, native profiling: CPU instruction count under payload scaling at $n=30$. The fixed cryptographic cost of Coded MBRB is high, but full-payload processing makes Bracha and AFRT grow with payload size.}
\label{fig:app-cpu-vs-payload}
\end{figure}

\paragraph*{Native profiling: CPU instructions under payload scaling.}
Fig.~\ref{fig:app-cpu-vs-payload} reports CPU instructions at $n=30$ while payload size grows from 100~KB to 8~MB. Coded MBRB has a high fixed computational cost, while Bracha and AFRT grow with payload size because they process full-payload messages repeatedly. Bracha overtakes AFRT around 500~KB and overtakes Coded MBRB around 6~MB. This result shows that the CPU ranking depends on payload size, not only on algorithm class.

\begin{figure}
\centering
\includegraphics[width=0.78\linewidth]{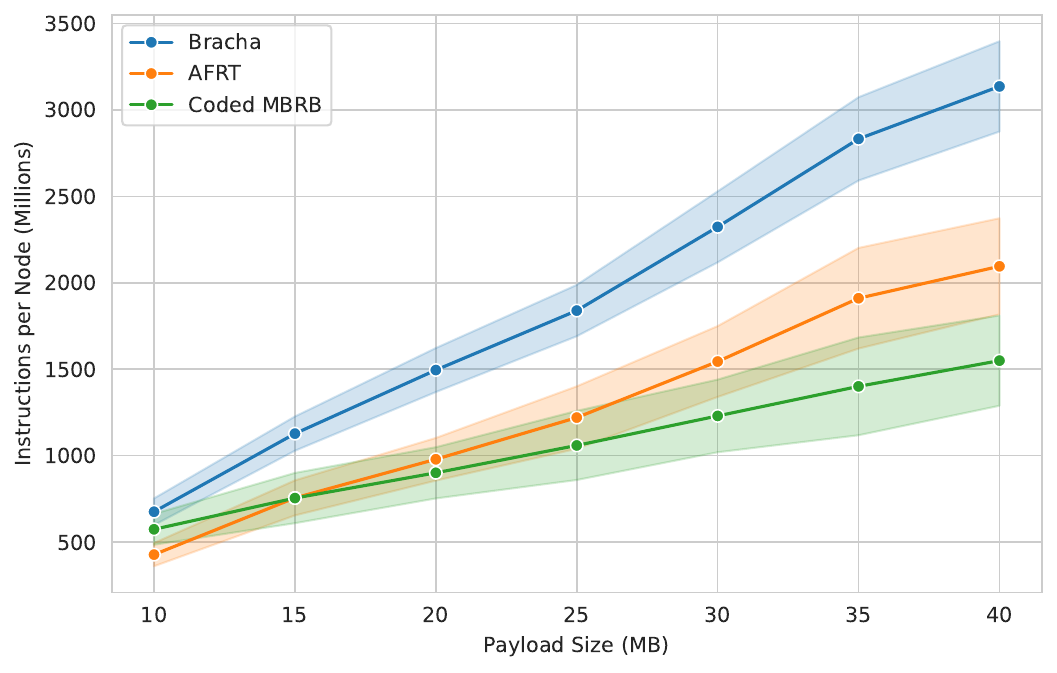}
\caption{Exp-1, native profiling: extended CPU instruction count under payload scaling at $n=10$. The extended test is run at $n=10$ because larger payloads exhausted the host memory at $n=30$.}
\label{fig:app-cpu-vs-payload-extended}
\end{figure}

\paragraph*{Native profiling: extended CPU payload scaling.}
Fig.~\ref{fig:app-cpu-vs-payload-extended} reports the extended CPU experiment from 10~MB to 40~MB at $n=10$. The 30-node configuration exhausted the 32~GB host memory at these larger payloads, so the extended test uses fewer nodes. The result shows AFRT surpassing Coded MBRB around a 15~MB payload. This strengthens the RQ2 conclusion that cryptographic fixed cost is not the only relevant CPU cost: repeated allocation, copying, hashing, and serialization of large payloads can dominate as payloads increase.

\begin{figure}
\centering
\includegraphics[width=0.78\linewidth]{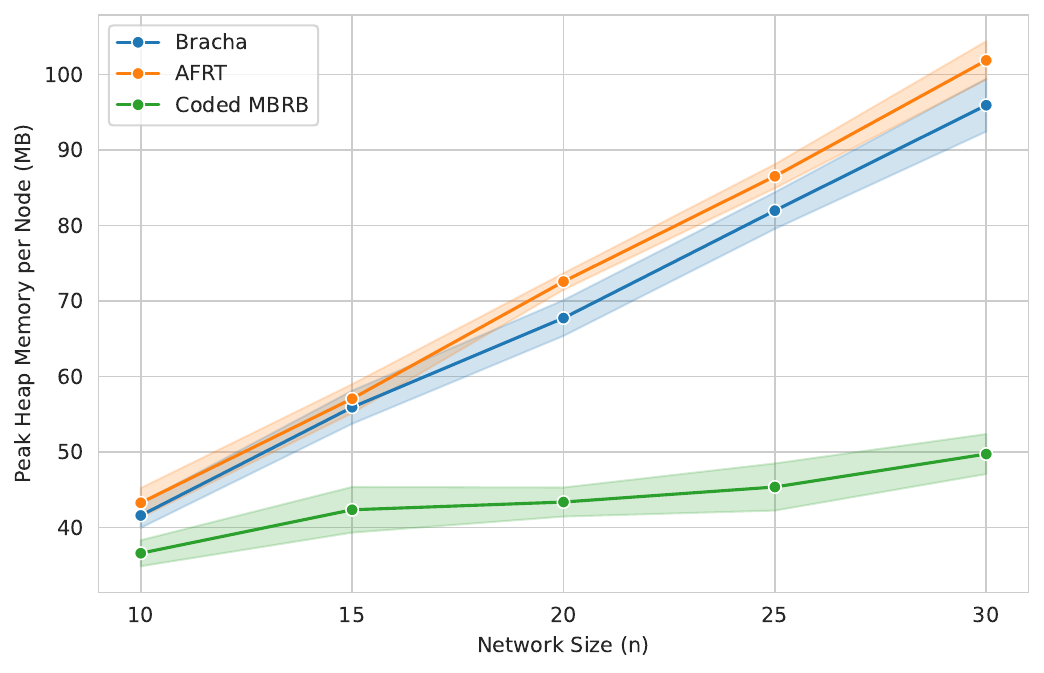}
\caption{Exp-1, native profiling: peak heap under node scaling with a 1~MB payload. Coded MBRB grows more slowly in peak heap than Bracha and AFRT.}
\label{fig:app-peak-heap-vs-nodes}
\end{figure}

\paragraph*{Native profiling: peak heap under node scaling.}
Fig.~\ref{fig:app-peak-heap-vs-nodes} reports peak heap as the number of nodes grows with a fixed 1~MB payload. All algorithms start near 40~MB at $n=10$. At $n=30$, Bracha and AFRT approach 100~MB, while Coded MBRB remains close to 50~MB. This result indicates that the memory advantage of fragment-based dissemination is visible even at a 1~MB payload.

\begin{figure}
\centering
\begin{subfigure}{0.48\textwidth}
\centering
\includegraphics[width=\textwidth]{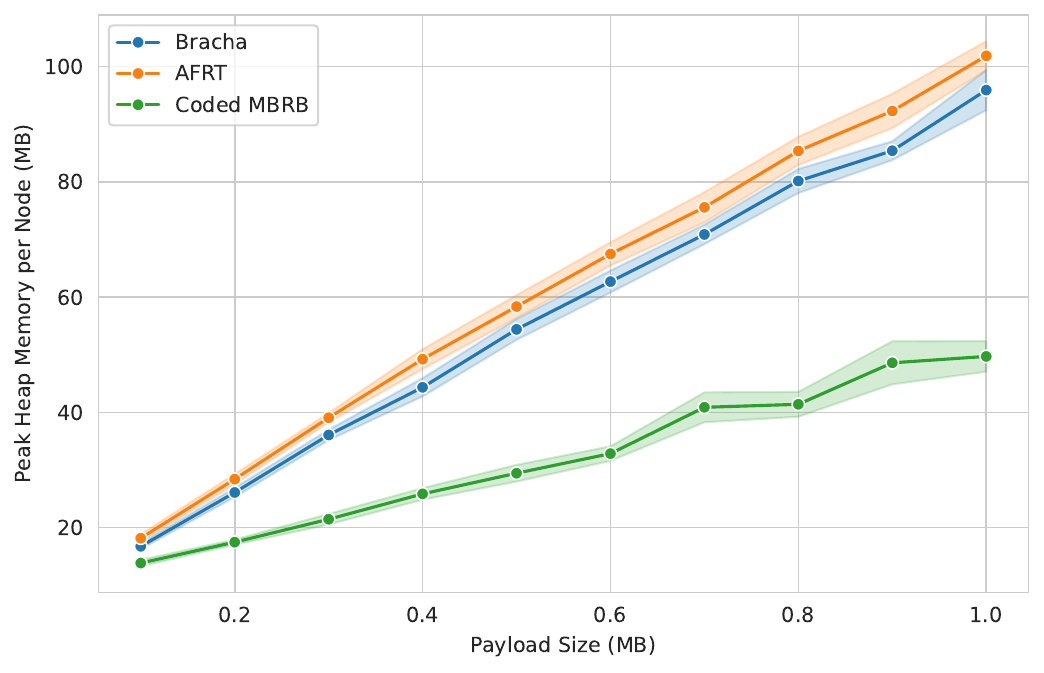}
\caption{100~KB--1~MB}
\end{subfigure}
\hfill
\begin{subfigure}{0.48\textwidth}
\centering
\includegraphics[width=\textwidth]{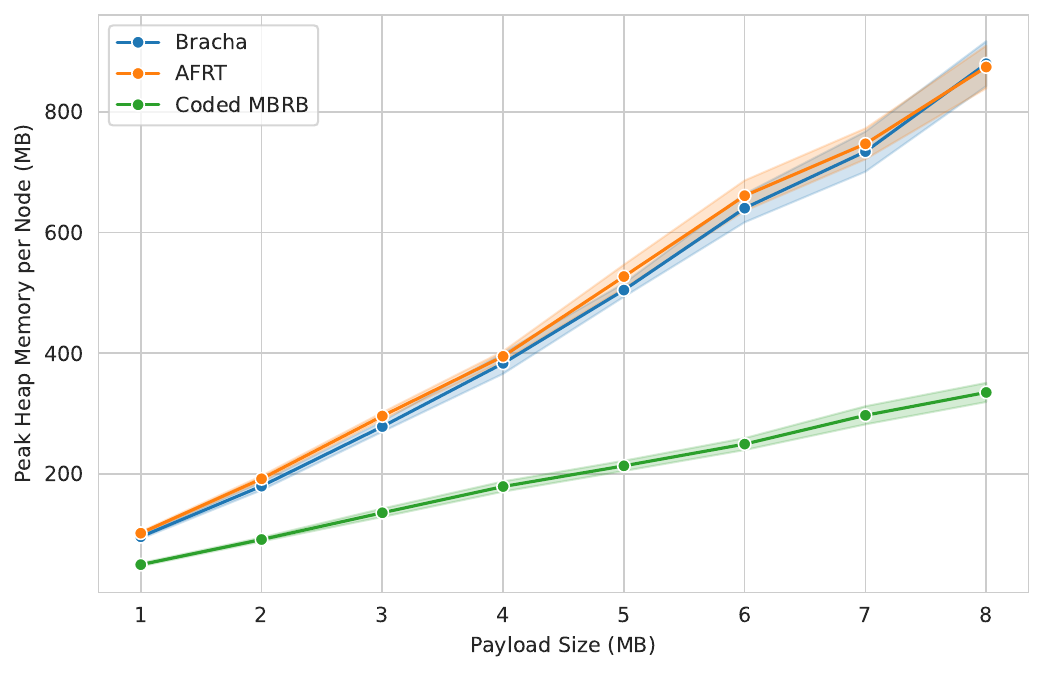}
\caption{1--8~MB}
\end{subfigure}
\caption{Exp-1, native profiling: peak heap under payload scaling at $n=30$. Coded MBRB uses substantially less peak heap for larger payloads.}
\label{fig:app-peak-heap-vs-payload}
\end{figure}

\paragraph*{Native profiling: peak heap under payload scaling.}
Fig.~\ref{fig:app-peak-heap-vs-payload} reports peak heap at $n=30$ as the payload size grows. At an 8~MB payload, Bracha and AFRT exceed 800~MB peak heap, while Coded MBRB is approximately 350~MB. 

\begin{figure}
\centering
\includegraphics[width=0.78\linewidth]{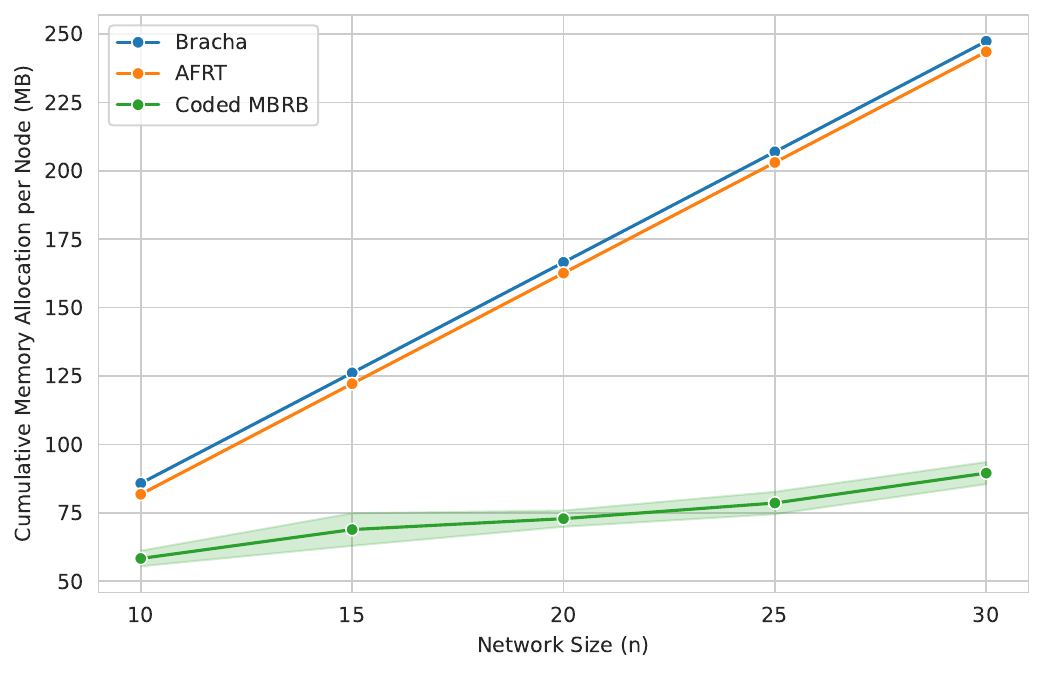}
\caption{Exp-1, native profiling: cumulative allocation under node scaling with a 1~MB payload. Cumulative allocation captures allocation churn caused by repeated full-payload handling.}
\label{fig:app-cumulative-alloc-vs-nodes}
\end{figure}

\paragraph*{Native profiling: cumulative allocation under node scaling.}
Fig.~\ref{fig:app-cumulative-alloc-vs-nodes} reports cumulative allocation as $n$ grows. We additionally report cumulative allocation because it captures allocation churn that is not fully visible from peak heap alone. Bracha and AFRT show substantially larger cumulative allocation than Coded MBRB under node scaling.

\begin{figure}
\centering
\begin{subfigure}{0.48\textwidth}
\centering
\includegraphics[width=\textwidth]{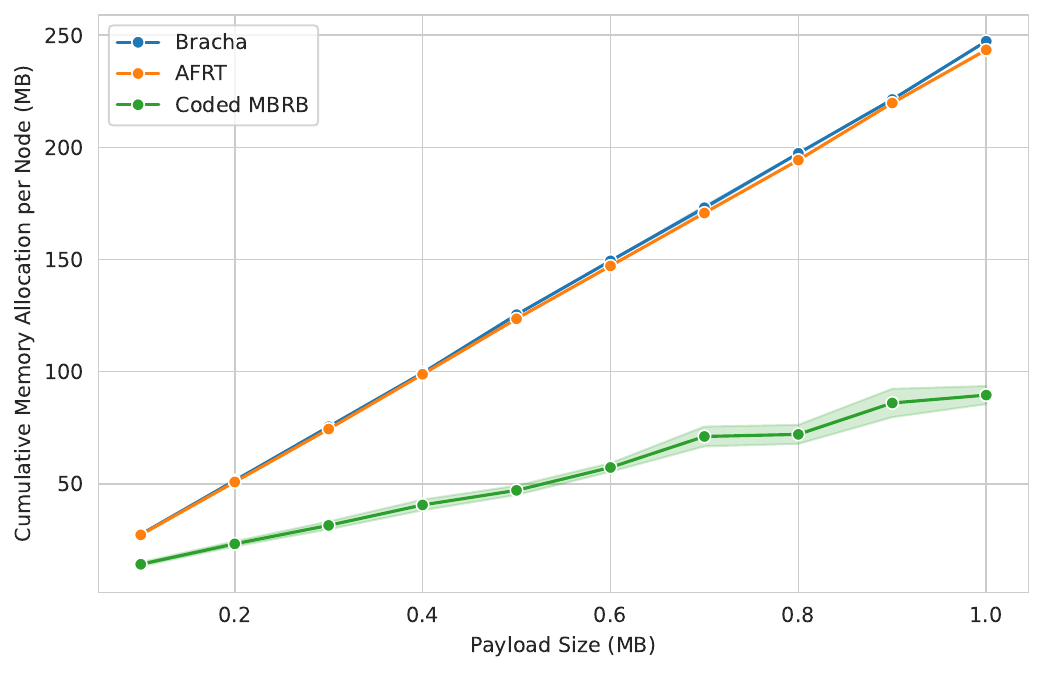}
\caption{100~KB--1~MB}
\end{subfigure}
\hfill
\begin{subfigure}{0.48\textwidth}
\centering
\includegraphics[width=\textwidth]{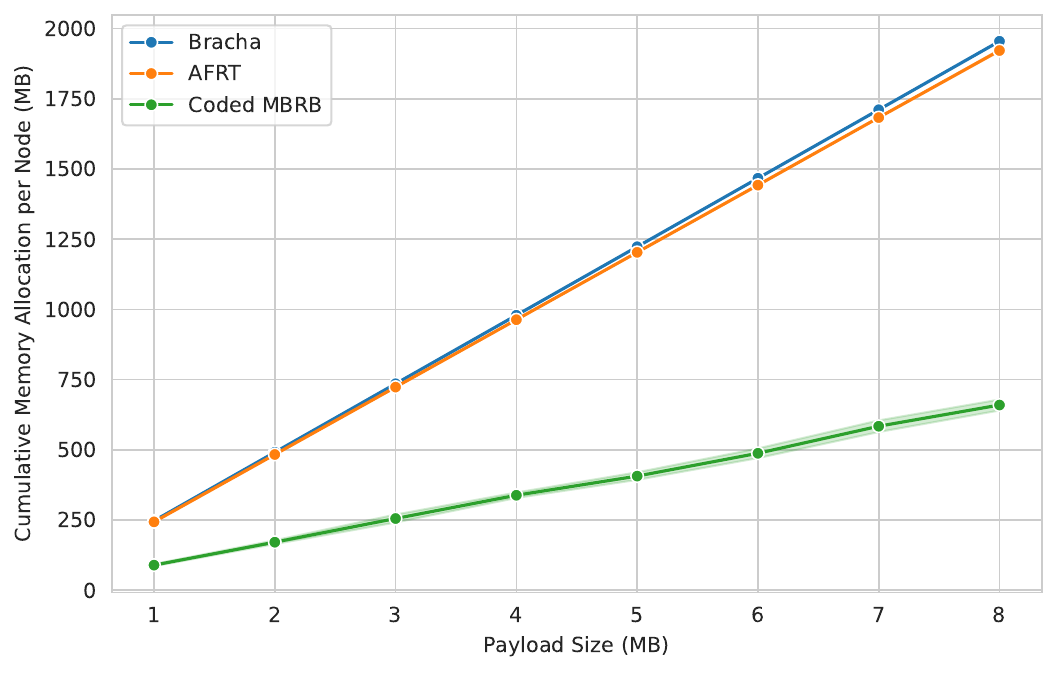}
\caption{1--8~MB}
\end{subfigure}
\caption{Exp-1, native profiling: cumulative allocation under payload scaling at $n=30$. Bracha and AFRT allocate substantially more memory as payload size grows.}
\label{fig:app-cumulative-alloc-vs-payload}
\end{figure}

\paragraph*{Native profiling: cumulative allocation under payload scaling.}
Fig.~\ref{fig:app-cumulative-alloc-vs-payload} reports cumulative allocation under payload scaling at $n=30$. Bracha and AFRT increase sharply as payload size grows, while Coded MBRB remains much lower. This result supports the interpretation that full-payload dissemination affects not only network traffic but also allocation behavior.

\subsection{Exp-2: Deployment Latency}
\label{sec:experiment2-results}

Experiment~2 answers RQ3. GCP is the primary deployment-oriented latency environment. FABRIC is retained as supplementary evidence because it exercises a more distributed infrastructure but requires RTT and site-placement context for careful interpretation.

\Subsubsection{Google Cloud Platform Results}
\label{sec:gcp-results}

\begin{figure}[ht]
\centering
\includegraphics[width=0.78\linewidth]{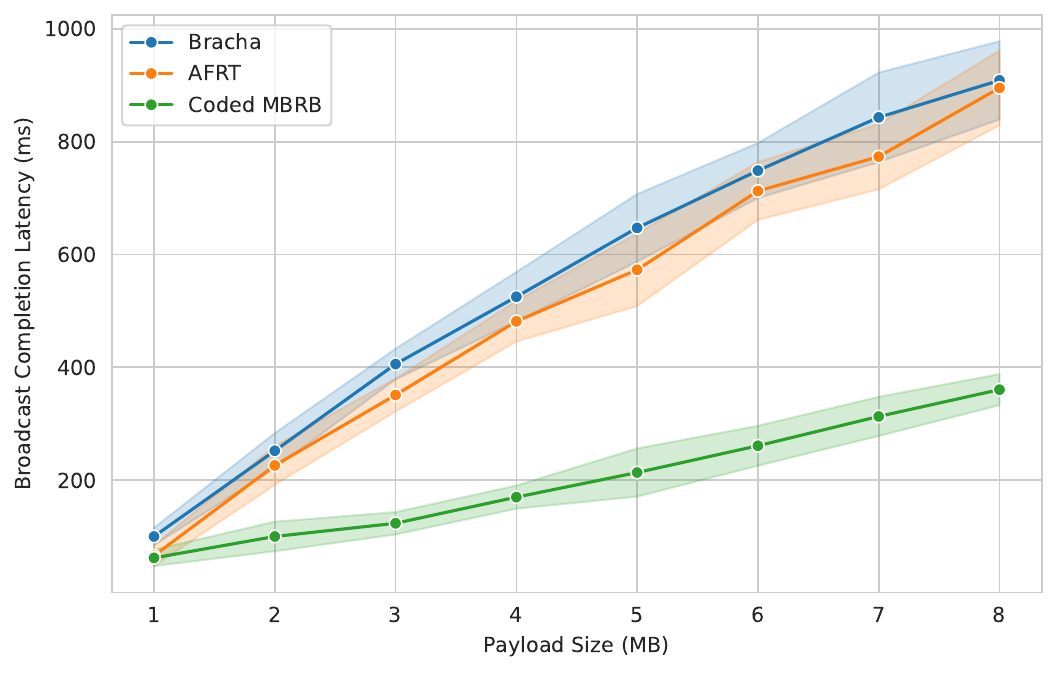}
\caption{Experiment~2, GCP: broadcast completion latency under payload scaling at $n=5$. Coded MBRB remains below 400~ms at 8~MB, while Bracha and AFRT reach approximately 900~ms.}
\label{fig:app-gcp-latency-vs-payload}
\end{figure}

\paragraph*{GCP: payload scaling.}
Fig.~\ref{fig:app-gcp-latency-vs-payload} reports broadcast completion latency at $n=5$ as payload size grows to 8~MB. Bracha and AFRT reach approximately 900~ms at 8~MB, while Coded MBRB remains below 400~ms. This result shows that the communication savings observed in Shadow can translate into lower completion latency in the evaluated cloud deployment.

\begin{figure}
\centering
\includegraphics[width=0.78\linewidth]{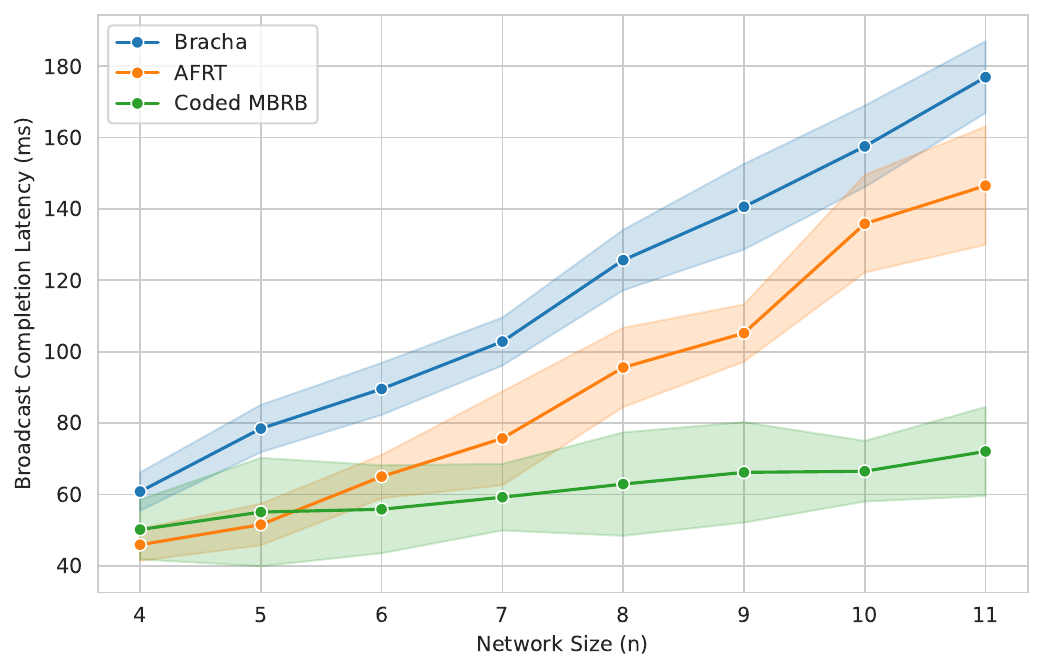}
\caption{Experiment~2, GCP: broadcast completion latency under node scaling with a 1~MB payload. Coded MBRB has the lowest latency at $n=11$ in the evaluated cloud setting.}
\label{fig:app-gcp-latency-vs-nodes}
\end{figure}

\paragraph*{GCP: node scaling.}
Fig.~\ref{fig:app-gcp-latency-vs-nodes} reports completion latency from $n=4$ to $n=11$ with a fixed 1~MB payload. At $n=11$, Coded MBRB is approximately 70~ms, compared with about 180~ms for Bracha and nearly 150~ms for AFRT. Although this is a small cloud deployment, it gives deployment-oriented evidence that Coded MBRB's lower byte volume can outweigh its additional local cryptographic work.

\Subsubsection{FABRIC Results}
\label{sec:fabric-results}

\paragraph*{FABRIC: interpretation context.}
The FABRIC results are supplementary deployment evidence. They are included because they exercise a wider distributed infrastructure and larger configurations than GCP. However, the latency curve for node scaling exhibits a noticeable fluctuation around $n = 15$. Based on our RTT measurements, where inter-site RTT reach up to 81.00~ms compared to sub-millisecond intra-site RTT, this variance reflects the placement of nodes across distant geographic sites rather than a bottleneck in the algorithms themselves. Because of this environmental variability, we treat the FABRIC data as supplementary evidence and rely primarily on Shadow and GCP for our controlled scaling and cloud-latency conclusions.

\begin{figure}
\centering
\includegraphics[width=0.78\linewidth]{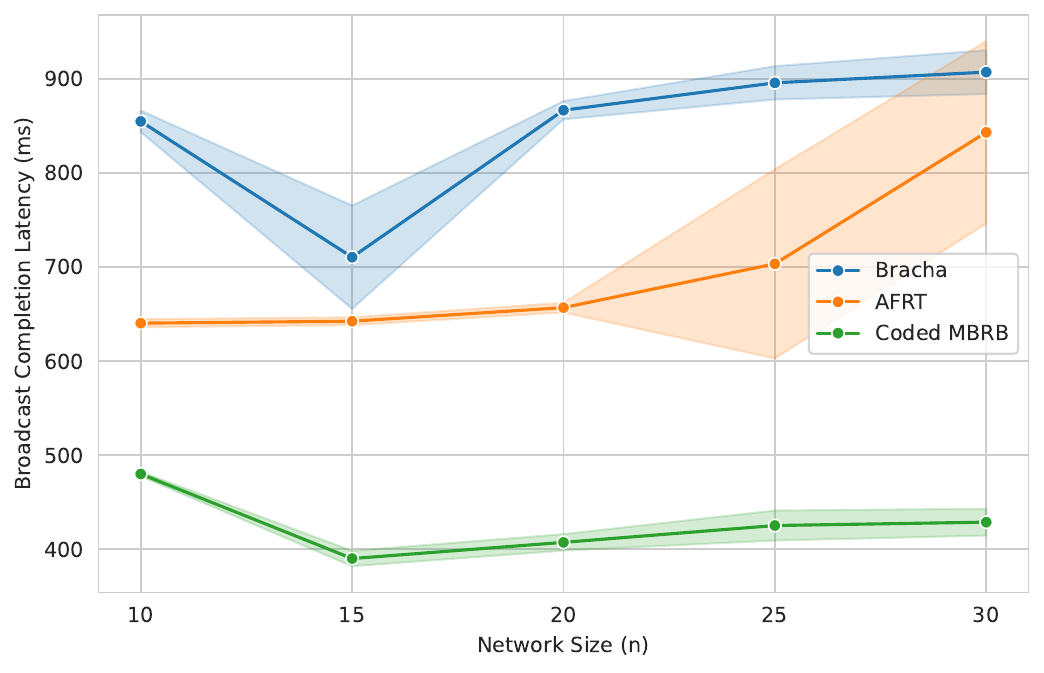}
\caption{Experiment~2, FABRIC: supplementary broadcast completion latency under node scaling with a 1~MB payload. This figure should be interpreted together with FABRIC RTT and site-placement measurements.}
\label{fig:app-fabric-latency-vs-nodes}
\end{figure}

\paragraph*{FABRIC: node scaling.}
Fig.~\ref{fig:app-fabric-latency-vs-nodes} reports completion latency under node scaling at a 1~MB payload, demonstrating the algorithmic behavior across a geographically distributed testbed.

\begin{figure}
\centering
\includegraphics[width=0.78\linewidth]{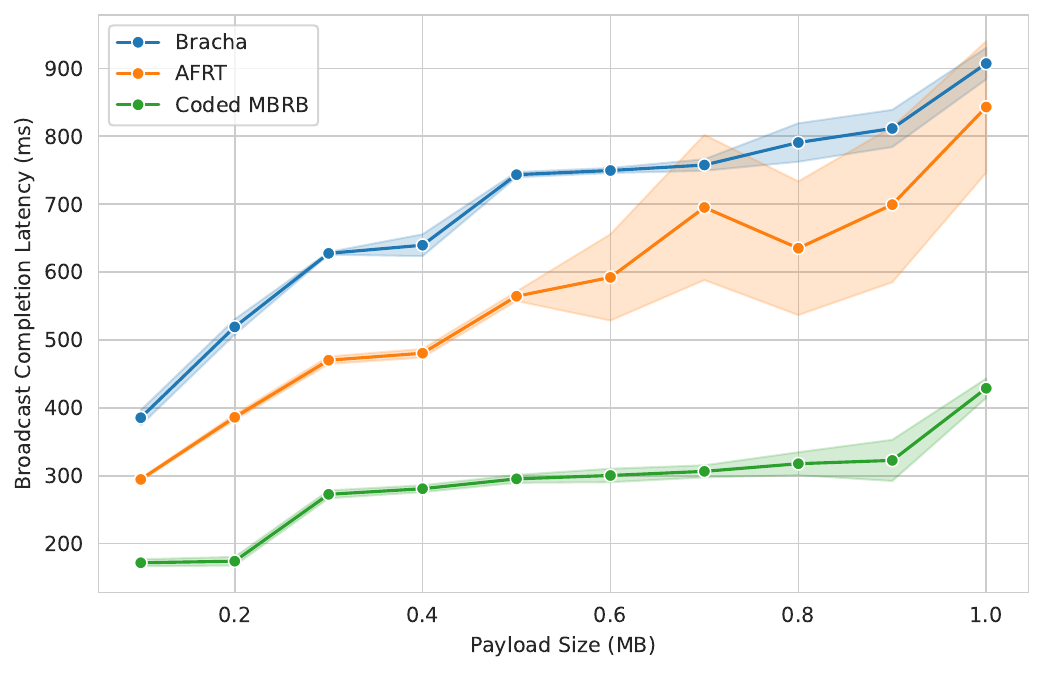}
\caption{Experiment~2, FABRIC: supplementary latency under micro-payload scaling at $n=30$.}
\label{fig:app-fabric-latency-vs-payload-micro}
\end{figure}

\paragraph*{FABRIC: micro-payload scaling.}
Fig.~\ref{fig:app-fabric-latency-vs-payload-micro} reports FABRIC latency under payload scaling from 100~KB to 1~MB at $n=30$. This range helps expose fixed overheads and small-payload behavior.

\begin{figure}
\centering
\includegraphics[width=0.78\linewidth]{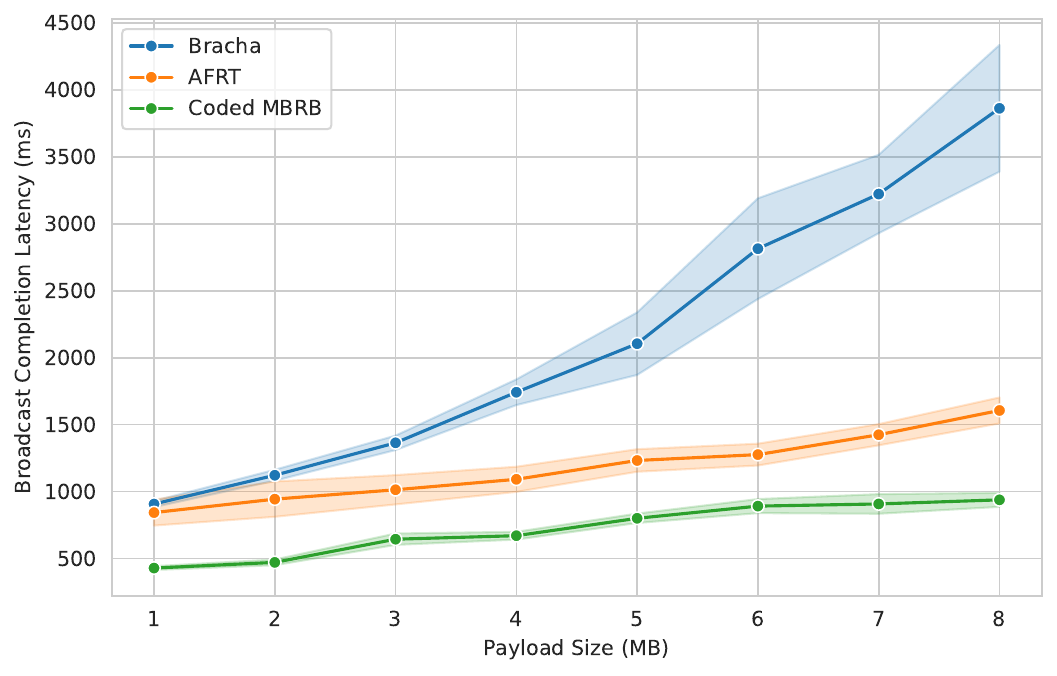}
\caption{Experiment~2, FABRIC: supplementary latency under macro-payload scaling at $n=30$.}
\label{fig:app-fabric-latency-vs-payload-macro}
\end{figure}

\paragraph*{FABRIC: macro-payload scaling.}
Fig.~\ref{fig:app-fabric-latency-vs-payload-macro} reports FABRIC latency from 1~MB to 8~MB at $n=30$. This range corresponds to the main macro-payload scaling interval used elsewhere in the evaluation.

\begin{figure}
\centering
\includegraphics[width=0.78\linewidth]{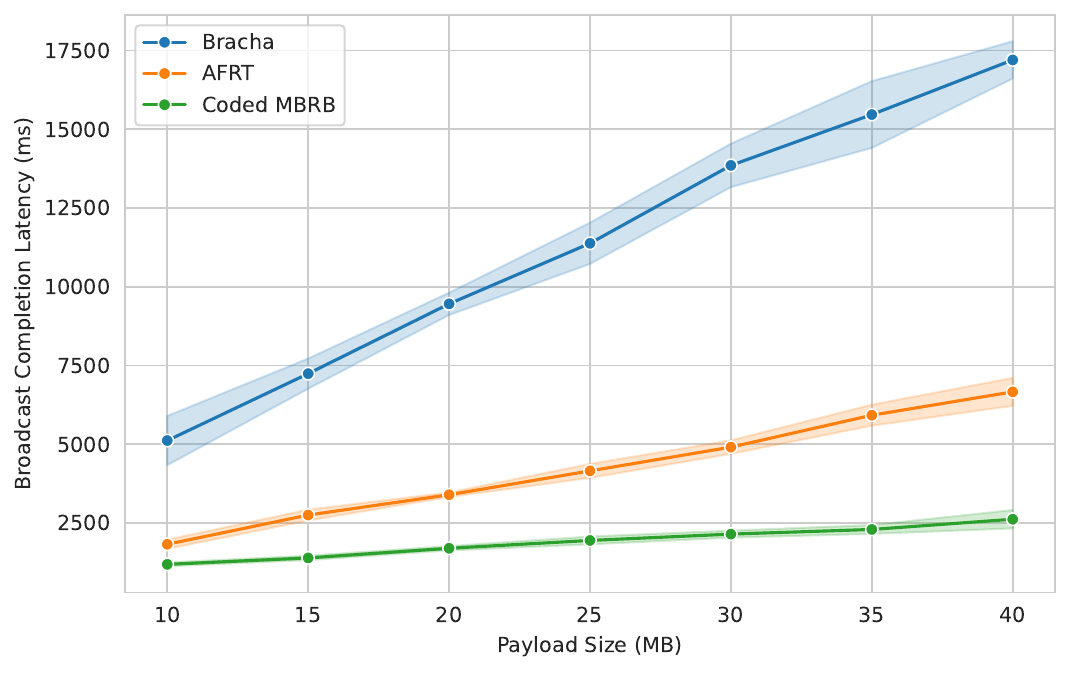}
\caption{Experiment~2, FABRIC: supplementary latency under extended payload scaling at $n=30$.}
\label{fig:app-fabric-latency-vs-payload-extended}
\end{figure}

\paragraph*{FABRIC: extended payload scaling.}
Fig.~\ref{fig:app-fabric-latency-vs-payload-extended} reports FABRIC latency from 10~MB to 40~MB at $n=30$, testing deployment latency under the geographic network conditions detailed previously.

\paragraph*{FABRIC RTT measurements.}
The RTT distribution among five FABRIC sites (TACC, UTAH, NCSA, MAX, and MICH) was measured immediately before every single broadcast execution during the initial startup handshake. Across the approximately 3 hours and 50 minutes it took to execute all performed algorithm tests on FABRIC, these measurements resulted in 690,900 individual RTT data points. Across all configurations, the minimum observed RTT was 0.06~ms, the median was 31.24~ms, the mean was 30.64~ms, and the maximum was 81.00~ms. Isolating the data by site reveals 118,500 intra-site pairs averaging 0.19~ms and 572,400 inter-site pairs averaging 36.95~ms. This results in high variance of 375.21~ms$^2$. 

\subsection{Exp-3: Fault Injection, Threshold Behavior, and Parser Checks}
\label{sec:experiment3-results}

Experiment~3 answers RQ4. The result items in this subsection are from Shadow simulation, native profiling and parser post-processing.

\begin{figure}[ht]
\centering
\includegraphics[width=0.78\linewidth]{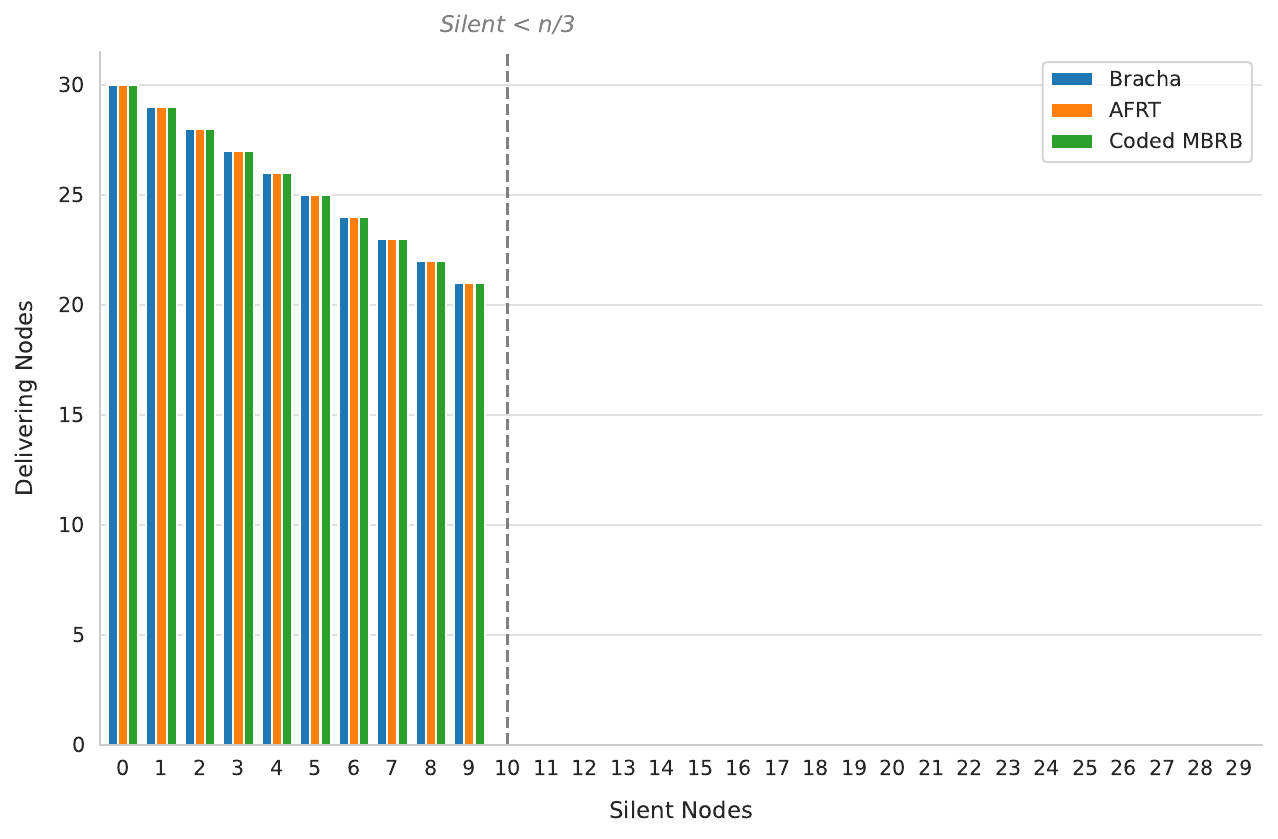}
\caption{Exp-3, Shadow: delivering nodes under silent-node scaling at $n=30$ and a 1~MB payload. Delivery follows the expected threshold behavior and drops to zero when the configured threshold is exceeded.}
\label{fig:app-delivering-vs-silent}
\end{figure}

\paragraph*{Shadow: silent-node threshold behavior.}
Fig.~\ref{fig:app-delivering-vs-silent} reports the number of delivering nodes as the number of silent nodes increases at $n=30$ and a 1~MB payload. For Bracha, the threshold condition $n>3t$ permits $t=9$ faults at $n=30$, leaving 21 non-silent nodes. At 9 silent nodes, the observed number of delivering nodes is 21. At 10 silent nodes, the threshold is exceeded and the observed number of delivering nodes drops to zero. AFRT and Coded MBRB follow the corresponding threshold behavior for their configured $t$ and $d$ values.

\begin{figure}
\centering
\includegraphics[width=0.78\linewidth]{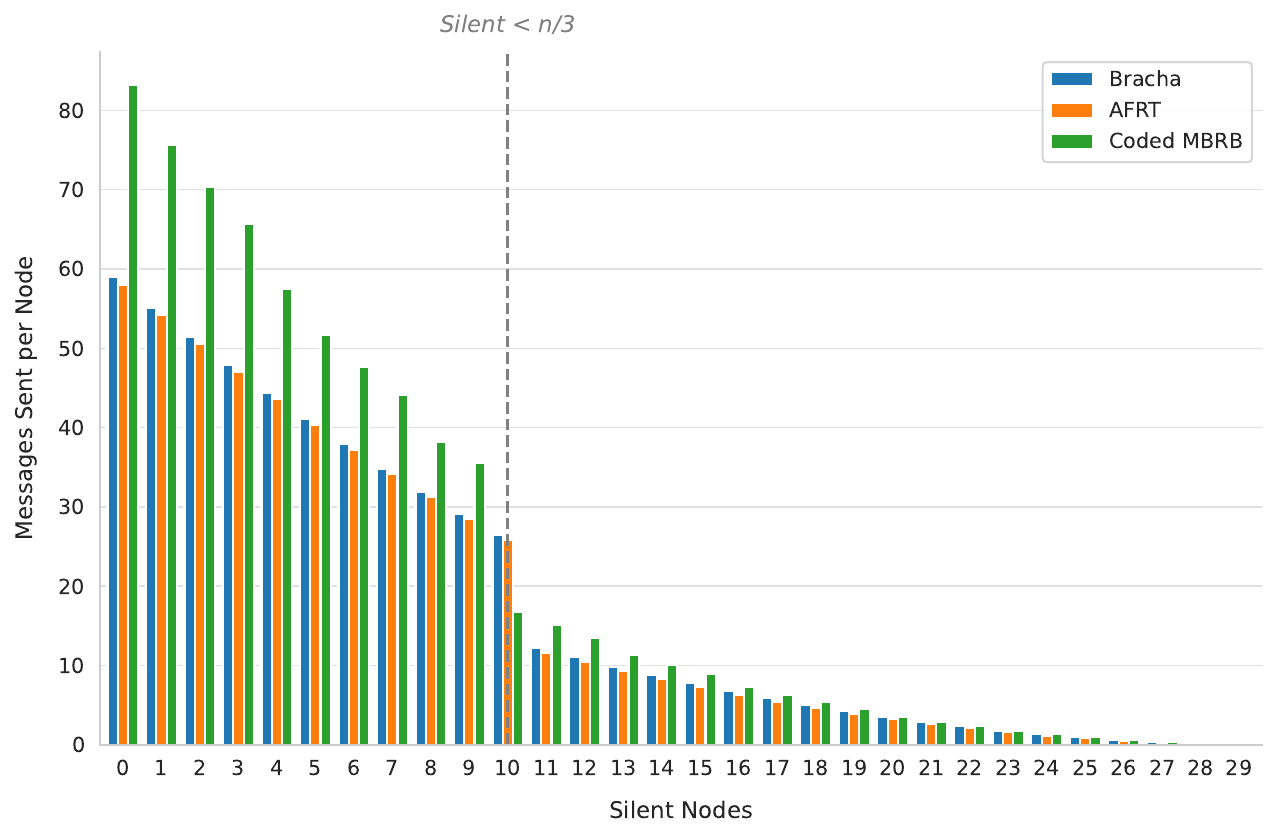}
\caption{Exp-3, Shadow: average messages per node under silent-node scaling at $n=30$ and a 1~MB payload. Message volume decreases as silent nodes suppress participation.}
\label{fig:app-messages-vs-silent-bar}
\end{figure}

\paragraph*{Shadow: average messages under silent nodes.}
Fig.~\ref{fig:app-messages-vs-silent-bar} reports the average messages sent per node as the number of silent nodes increases. Message volume decreases as fewer active nodes participate in the broadcast. This supports the performance-degradation interpretation: injected silence reduces the amount of work generated by the algorithm rather than producing unexpected additional message storms.

\begin{figure}
\centering
\includegraphics[width=0.78\linewidth]{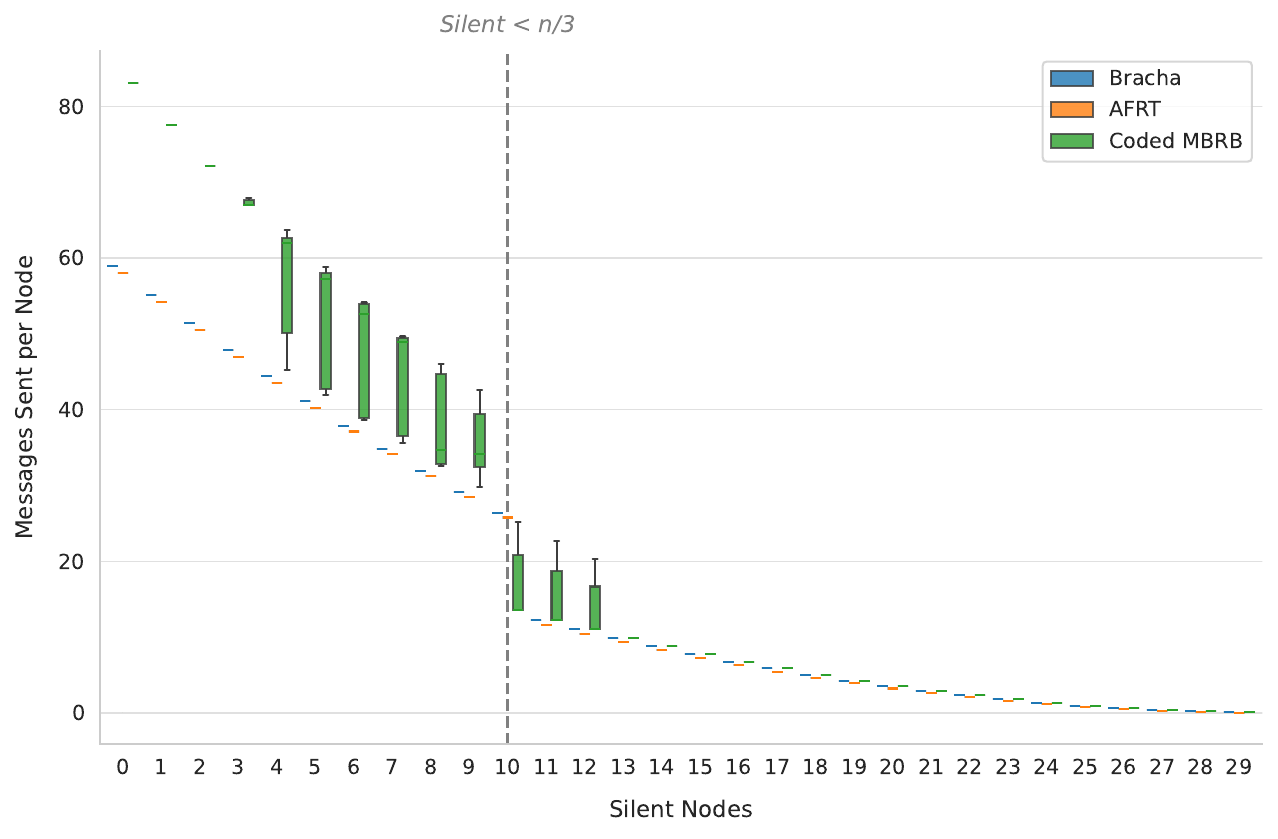}
\caption{Exp-3, Shadow: distribution of messages per node under silent-node scaling. The distribution complements the average-message plot by showing the spread across nodes.}
\label{fig:app-messages-vs-silent-box}
\end{figure}

\paragraph*{Shadow: message-distribution under silent nodes.}
Fig.~\ref{fig:app-messages-vs-silent-box} reports the distribution of messages sent per node. The distribution view shows whether work is balanced across the remaining active nodes or concentrated on selected nodes. In the tested configurations, the observed degradation is consistent with the reduction in active participants.

\begin{table}
\centering
\caption{Exp-3, Shadow: consistency checks across selected fault scenarios at $n=30$. 
}
\label{tab:fault-outcomes}
\label{tab:app-correctness-validation}
\scriptsize
\setlength{\tabcolsep}{4pt}
\renewcommand{\arraystretch}{1.2}
\begin{tabularx}{\textwidth}{
  >{\hsize=0.8\hsize\raggedright\arraybackslash}X
  >{\hsize=1.2\hsize\raggedright\arraybackslash}X
  >{\hsize=1.0\hsize\raggedright\arraybackslash}X
  >{\hsize=0.75\hsize\centering\arraybackslash}X
  >{\hsize=0.6\hsize\centering\arraybackslash}X
  >{\hsize=0.8\hsize\raggedright\arraybackslash}X
}
\toprule
\textbf{Algorithm} & \textbf{Fault scenario} & \textbf{Theoretical status} & \textbf{Delivery guarantee} & \textbf{Observed delivery} & \textbf{Detected violation} \\
\midrule
Bracha & \texttt{numSilent}=9, \texttt{numMsgDrops}=0 & Within $n>3t$ & 21 & 21 & No \\
Bracha & \texttt{numSilent}=10, \texttt{numMsgDrops}=0 & Threshold exceeded & No guarantee & 0 & No \\
AFRT & \texttt{numSilent}=5, \texttt{numMsgDrops}=7 & Within $n>3t+2d$ & 18 & 18 & No \\
AFRT & \texttt{numSilent}=7, \texttt{numMsgDrops}=5 & Threshold exceeded & No guarantee & 0 & No \\
Coded MBRB & \texttt{numSilent}=5, \texttt{numMsgDrops}=7 & Within $n>3t+2d$ & 18 & 18 & No \\
Coded MBRB & \texttt{numSilent}=7, \texttt{numMsgDrops}=5 & Threshold exceeded & No guarantee & 0 & No \\
All algorithms & \texttt{twinsSender=true} & Partition-based equivocation & No guarantee & 0 & No \\
\bottomrule
\end{tabularx}
\end{table}

\paragraph*{Shadow: selected fault-scenario outcomes.}
{Table~\ref{tab:app-correctness-validation} summarizes the selected threshold and equivocation scenarios at $n=30$. Within the sufficient bounds, the delivery guarantee matches the observed count: 21 for Bracha and 18 for both AFRT and Coded MBRB. Beyond these bounds, and under the partition-based equivocation scenario, no delivery guarantee applies; zero deliveries were observed.}
These sufficient bounds are not claimed to be sharp empirical cutoffs. Our experiments evaluate representative configurations within and immediately beyond the proven thresholds, but do not systematically characterize delivery behavior outside the guaranteed region.

\begin{figure}[ht]
\centering
\begin{subfigure}[t]{0.45\linewidth}
\centering
\includegraphics[width=\linewidth]{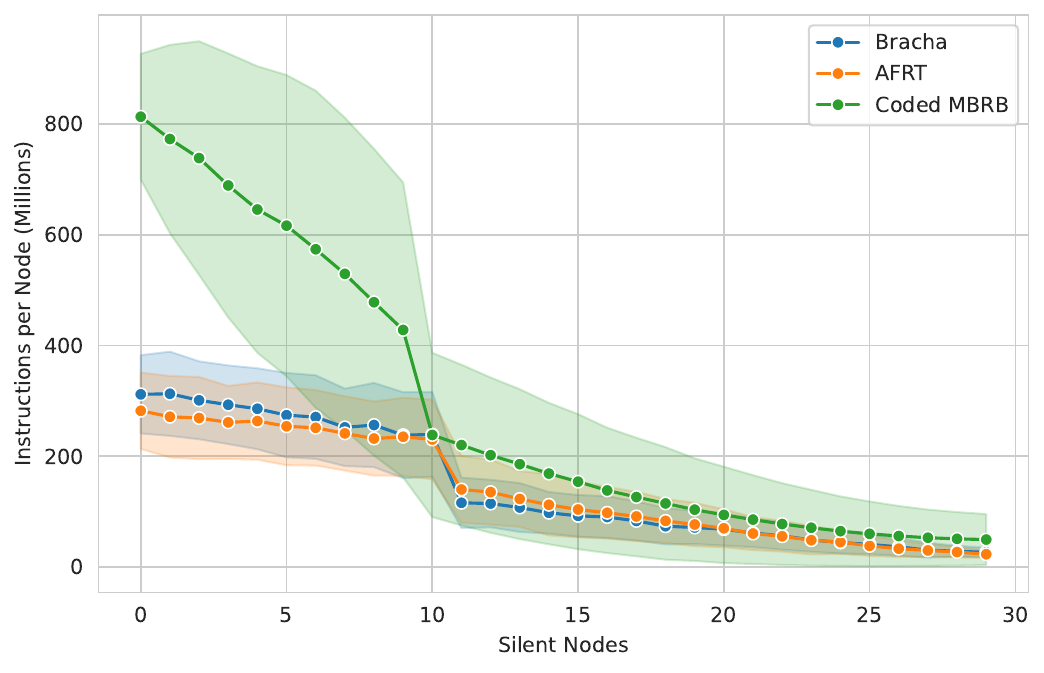}
\caption{CPU instructions.}
\label{fig:local_cpu_instructions_vs_silent_payload_size_1MB_0-29}
\end{subfigure}

\hfill
\begin{subfigure}[t]{0.45\linewidth}
\centering
\includegraphics[width=\linewidth]{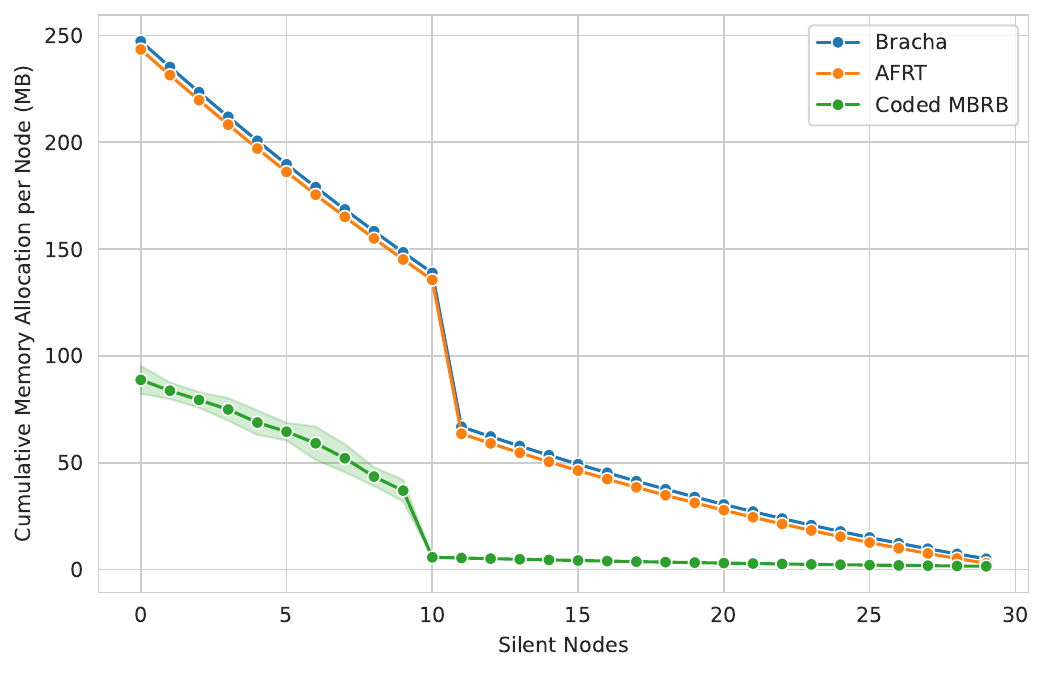}
\caption{Cumulative allocation.}
\label{fig:local_cumulative_alloc_vs_silent_payload_size_1MB_0-29}
\end{subfigure}
\hfill
\begin{subfigure}[t]{0.45\linewidth}
\centering
\includegraphics[width=\linewidth]{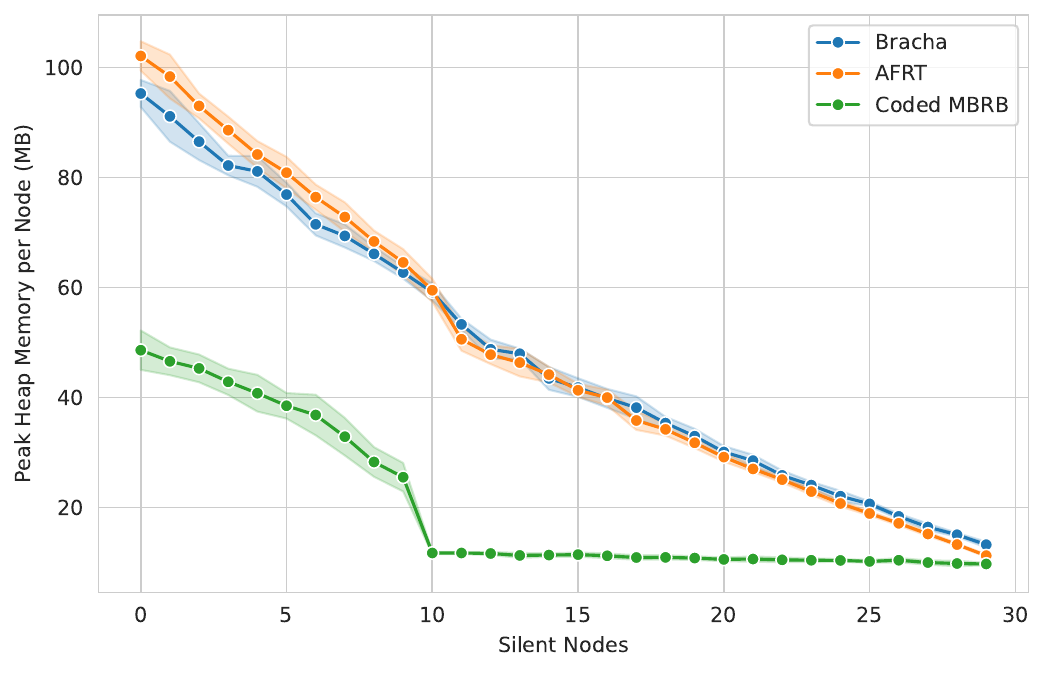}
\caption{Peak heap.}
\label{fig:local_peak_heap_vs_silent_payload_size_1MB_0-29}
\end{subfigure}

\caption{Exp-3, native profiling: local resource usage under silent-node scaling at a 1~MB payload. The panels report CPU instruction count, cumulative allocation, and peak heap usage.}
\label{fig:local_resources_vs_silent_payload_size_1MB_0-29}
\end{figure}

\paragraph*{Native profiling: performance degradation.}
The fault-injection results also show predictable performance degradation. As the number of silent nodes grows, message volume, total transmitted data, CPU instructions, and memory allocation decrease because fewer nodes actively participate in the message exchange. At the maximum within-threshold silent-node configuration, average CPU instructions decrease by approximately 30\% for Bracha and AFRT, and around 50\% for Coded MBRB, peak heap usage decreases by 40–50\%, and cumulative memory allocation by 45–60\%. When the fault threshold is exceeded, local resource usage drops sharply because the protocols fail to reach required quorums and bypass execution phases. 
Figure~\ref{fig:local_resources_vs_silent_payload_size_1MB_0-29} compares the CPU instructions, peak heap usage, and cumulative memory allocation under silent-node faults. This result does not change the main performance ranking; rather, it shows that the injected silent-node fault reduces work roughly in proportion to the number of active participants.

\subsection{Additional Parser-Check Campaign}
\label{sec:parser-check-campaign}

The parser-check campaign is a key part of the reproducibility evidence. It checks the raw outputs produced by both the plotted experiments and additional fault-injection configurations. The parser checks duplicate delivery, conflicting delivery, validity against the sender payload when applicable, and spurious delivery.

The primary plotted results contribute 43,500 checked entries from 1,500 Shadow runs, 19,500 checked entries from 840 native profiling runs, 3,000 checked entries from 480 GCP deployment runs, 24,600 checked entries from 870 FABRIC deployment runs, and 27,000 checked entries from 900 native profiling fault-injection runs. 
{Additional Shadow tests cover $n\in\{10,20,30\}$, a 100~KB payload, all relevant silent-node and message-drop permutations under maximal-$t$ configurations with $d=0$ and maximal-$d$ configurations with $t=0$, cases with and without the equivocating sender, and the 10 recorded seeds.} 
These additional tests contribute 2,244,000 checked entries over 87,600 runs.

Across the full campaign, the parser checked 2,361,600 entries over 92,190 runs. It found zero duplicate-delivery, conflicting-delivery, invalid-delivery, or spurious-delivery violations in the tested configurations. This is implementation-level evidence over the generated executions. It is not a formal proof and it is not exhaustive Byzantine testing.

\subsection{Answers to the Research Questions}
\label{sec:extended-rq-answers}

\paragraph*{Answer to RQ1: Communication scalability.}
{Compared with AFRT, the directly comparable MBRB baseline, Coded MBRB substantially improves byte scalability but not message-count scalability.} In Shadow at $n=30$ with a 1~MB payload, Bracha and AFRT transmit ca 1.7~GB, whereas Coded MBRB transmits close to 0.1~GB (Fig.~\ref{fig:app-sim-data-sent-vs-nodes}). Under the same configuration, Coded MBRB sends roughly 2500 implementation messages, compared with ca 1750 for Bracha and AFRT (Fig.~\ref{fig:app-sim-messages-vs-nodes}). Under payload scaling at $n=30$, Bracha and AFRT reach ca 14~GB at an 8~MB payload, while Coded MBRB remains around 1~GB (Fig.~\ref{fig:app-sim-data-sent-vs-payload}). {The answer to RQ1 is therefore that Coded MBRB is far more scalable in transmitted bytes than AFRT, while Bracha serves as the classical BRB reference and both Bracha and AFRT use fewer implementation messages.}

\paragraph*{Answer to RQ2: Local resource cost.}
{Compared with AFRT, Coded MBRB pays a higher local computational cost at smaller payloads, but avoids the large-payload CPU and memory growth caused by full-payload dissemination.} In native profiling at $n=30$ and a 1~MB payload, Coded MBRB executes ca 800 million CPU instructions per node, compared with roughly 250 million for Bracha and AFRT (Fig.~\ref{fig:app-cpu-vs-nodes}). As payload size grows, Bracha overtakes Coded MBRB in CPU cost around 6~MB at $n=30$, and AFRT overtakes Coded MBRB around 15~MB in the extended $n=10$ experiment (Figures~\ref{fig:app-cpu-vs-payload} and~\ref{fig:app-cpu-vs-payload-extended}). Memory shows the same qualitative shift: at an 8~MB payload and $n=30$, Bracha and AFRT exceed 800~MB peak heap per node, while Coded MBRB remains around 350~MB (Fig.~\ref{fig:app-peak-heap-vs-payload}). 
{Thus, relative to AFRT, Coded MBRB becomes preferable for memory pressure and eventually for CPU cost in sufficiently large-payload regimes, although it remains more expensive in CPU at smaller payloads.}

\paragraph*{Answer to RQ3: Deployment latency.}
The GCP deployment shows that Coded MBRB's communication savings can translate into lower completion latency. At an 8~MB payload with $n=5$, Bracha and AFRT reach ca 900~ms, whereas Coded MBRB remains below 400~ms (Fig.~\ref{fig:app-gcp-latency-vs-payload}). At $n=11$ with a 1~MB payload, Coded MBRB is roughly 70~ms, compared with about 180~ms for Bracha and nearly 150~ms for AFRT (Fig.~\ref{fig:app-gcp-latency-vs-nodes}). FABRIC provides additional deployment-oriented measurements up to $n=30$ and larger payloads. Because the FABRIC nodes are geographically distributed across five US sites with high inter-site RTT (averaging 36.95~ms), these results naturally exhibit higher baseline latencies and higher variance. However, they {show the same qualitative trend in the evaluated configurations: Coded MBRB achieves lower latency as payloads grow.}
The answer to RQ3 is affirmative in both deployment environments, GCP and FABRIC, where Coded MBRB achieved the lowest latency in our experiments.

\paragraph*{Answer to RQ4: Fault-injection behavior.}
The tested executions behaved consistently with the expected threshold behavior and produced no parser-detected specification violations.
In the $n=30$ silent-node experiment, {with 9 silent nodes, all 21 non-silent nodes are guaranteed to deliver and did so}, while 10 silent nodes exceed the threshold and yield zero deliveries (Fig.~\ref{fig:app-delivering-vs-silent} and Table~\ref{tab:fault-outcomes}). For AFRT and Coded MBRB with \texttt{numSilent}=5 and \texttt{numMsgDrops}=7, both the guaranteed delivery and the observed counts are 18. 
{In the threshold-exceeded and partition-based equivocation scenarios, no delivery guarantee applies, and zero deliveries were observed in the evaluated configurations.}
Across 92,190 runs and 2,361,600 parser-checked entries, the parser found zero duplicate-delivery, conflicting-delivery, invalid-delivery, or spurious-delivery violations. This answers RQ4 for the tested fault scenarios, while leaving exhaustive Byzantine testing outside the paper's scope.

\section{Conclusion}
\label{sec:conclusion}

We evaluated Bracha's BRB algorithm, AFRT, and Coded MBRB in a shared Go benchmark framework using Shadow simulation, native profiling, and deployment-oriented latency measurements. 
The results expose a concrete systems trade-off. Relative to AFRT, the directly comparable MBRB baseline, Coded MBRB reduces data movement and deployment latency for larger payloads, but shifts cost to cryptographic and coding computation. Bracha serves as the classical BRB reference and exhibits similar full-payload communication behavior. At $n=30$ with a 1~MB payload, Coded MBRB sends about $14\times$ less data than AFRT (and similarly than Bracha), but executes about $3\times$ more CPU instructions per node. At an 8~MB payload, Coded MBRB uses around 350~MB peak heap, whereas both AFRT and Bracha exceed 800~MB.
In the GCP deployment, Coded MBRB remains below 400~ms while AFRT and Bracha reach about 900~ms, and the supplementary FABRIC experiments show the same qualitative large-payload latency advantage over a broader distributed testbed.
Across 92,190 runs, {the parser found none of the checked violations in the tested configurations.} The artifact, found in our open-source repository~\cite{brb-eval-open-source-artifact}, supports extending the benchmark to additional algorithms, workloads, environments, and fault-injection semantics.

\bibliographystyle{splncs04}
\bibliography{references}

@inproceedings{DBLP:conf/podc/AbrahamN0X21,
  author       = {Ittai Abraham and
                  Kartik Nayak and
                  Ling Ren and
                  Zhuolun Xiang},
  editor       = {Avery Miller and
                  Keren Censor{-}Hillel and
                  Janne H. Korhonen},
  title        = {Good-case Latency of Byzantine Broadcast: a Complete Categorization},
  booktitle    = {{PODC} '21: {ACM} Symposium on Principles of Distributed Computing,
                  Virtual Event, Italy, July 26-30, 2021},
  pages        = {331--341},
  publisher    = {{ACM}},
  year         = {2021},
}

@inproceedings{DBLP:conf/podc/AlhaddadDD0VXZ22,
  author       = {Nicolas Alhaddad and
                  Sourav Das and
                  Sisi Duan and
                  Ling Ren and
                  Mayank Varia and
                  Zhuolun Xiang and
                  Haibin Zhang},
  editor       = {Alessia Milani and
                  Philipp Woelfel},
  title        = {Balanced Byzantine Reliable Broadcast with Near-Optimal Communication
                  and Improved Computation},
  booktitle    = {Symposium on Principles of Distributed Computing, {PODC}},
  pages        = {399--417},
  publisher    = {{ACM}},
  year         = {2022},
}

@inproceedings{DBLP:conf/ccs/DasX021,
  author       = {Sourav Das and
                  Zhuolun Xiang and
                  Ling Ren},
  editor       = {Yongdae Kim and
                  Jong Kim and
                  Giovanni Vigna and
                  Elaine Shi},
  title        = {Asynchronous Data Dissemination and its Applications},
  booktitle    = {{SIGSAC} Conference on Computer and Communications Security, {CCS}},
  pages        = {2705--2721},
  publisher    = {{ACM}},
  year         = {2021},
}

@inproceedings{DBLP:conf/opodis/Locher24,
  author       = {Thomas Locher},
  editor       = {Silvia Bonomi and
                  Letterio Galletta and
                  Etienne Rivi{\`{e}}re and
                  Valerio Schiavoni},
  title        = {{Byzantine} Reliable Broadcast with Low Communication and Time Complexity},
  booktitle    = {28th International Conference on Principles of Distributed Systems {OPODIS}},
  series       = {LIPIcs},
  volume       = {324},
  pages        = {16:1--16:17},
  publisher    = {Schloss Dagstuhl - Leibniz-Zentrum f{\"{u}}r Informatik},
  year         = {2024},
}

@inproceedings{DBLP:conf/opodis/Locher25,
  author       = {Thomas Locher},
  editor       = {Andrei Arusoaie and
                  Emanuel Onica and
                  Michael Spear and
                  Sara Tucci Piergiovanni},
  title        = {Efficient {Byzantine} Reliable Broadcast in the Failure Case},
  booktitle    = {29th International Conference on Principles of Distributed Systems {OPODIS}},
  series       = {LIPIcs},
  volume       = {361},
  pages        = {12:1--12:20},
  publisher    = {Schloss Dagstuhl - Leibniz-Zentrum f{\"{u}}r Informatik},
  year         = {2025},
}

@inproceedings{DBLP:conf/eurocrypt/LocherS25,
  author       = {Thomas Locher and
                  Victor Shoup},
  editor       = {Serge Fehr and
                  Pierre{-}Alain Fouque},
  title        = {MiniCast: Minimizing the Communication Complexity of Reliable Broadcast},
  booktitle    = {Advances in Cryptology - {EUROCRYPT} 2025 - 44th Annual International
                  Conference on the Theory and Applications of Cryptographic Techniques,
                  Madrid, Spain, May 4-8, 2025, Proceedings, Part {V}},
  series       = {Lecture Notes in Computer Science},
  volume       = {15605},
  pages        = {96--115},
  publisher    = {Springer},
  year         = {2025}
}

@article{DBLP:journals/iacr/LocherS25,
  author       = {Thomas Locher and
                  Victor Shoup},
  title        = {Improving the Round Complexity of MiniCast},
  journal      = {{IACR} Cryptol. ePrint Arch.},
  volume       = {2025},
  pages        = {779},
  year         = {2025},
}

@misc{DisatnikBoshoerCodedMBRB,
author       = {Benjamin Boshoer and
Yuval Disatnik},
title        = {{Coded MBRB} Implementation Repository},
howpublished = {\url{https://github.com/BenjaminBoshoer/CE-Final-Project}},
note         = {Bachelor's thesis implementation project, Bar-Ilan University},
year         = {2024},
urldate      = {2026-06-08}
}

@misc{brb-eval-open-source-artifact,
	author       = {Jesper Kullberg and Fabian Paglianno Persson},
	title        = {Evaluating {Byzantine} Reliable Broadcast Algorithms},
	year         = {2026},
	url          = {https://github.com/fabianPag/brb-eval},
}

@misc{AFRTRustImplementation,
	author = {Tom Picaud and
	Timothé Albouy},
	title = {{mbrb-rs}: Rust Implementation of Message-Adversary-Tolerant {Byzantine} Reliable Broadcast},
	howpublished = {\url{https://gitlab.inria.fr/WIDE/mbrb-rs/}},
	note = {Research implementation artifact},
	year = {2021},
	urldate = {2026-06-08}
}

@article{DBLP:journals/tcs/DuvignauRS23,
	author       = {Romaric Duvignau and
	Michel Raynal and
	Elad Michael Schiller},
	title        = {Self-stabilizing {Byzantine} fault-tolerant repeated reliable broadcast},
	journal      = {Theor. Comput. Sci.},
	volume       = {972},
	pages        = {114070},
	year         = {2023},
}

@inproceedings{DBLP:conf/wdag/CamaioniGMV22,
	author       = {Martina Camaioni and
	Rachid Guerraoui and
	Matteo Monti and
	Manuel Vidigueira},
	editor       = {Christian Scheideler},
	title        = {Oracular {Byzantine} Reliable Broadcast},
	booktitle    = {36th International Symposium on Distributed Computing, {DISC} 2022,
	Augusta, Georgia, USA, October 25-27, 2022},
	series       = {LIPIcs},
	pages        = {13:1--13:19},
	publisher    = {Schloss Dagstuhl - Leibniz-Zentrum f{\"{u}}r Informatik},
	year         = {2022},
}

@book{DBLP:books/sp/Raynal18,
	author       = {Michel Raynal},
	title        = {Fault-Tolerant Message-Passing Distributed Systems - An Algorithmic Approach},
	publisher    = {Springer},
	year         = {2018},
}

@inproceedings{DBLP:conf/opodis/BanoSCPLCM21,
	author       = {Shehar Bano and
	Alberto Sonnino and
	Andrey Chursin and
	Dmitri Perelman and
	Zekun Li and
	Avery Ching and
	Dahlia Malkhi},
	editor       = {Quentin Bramas and
	Vincent Gramoli and
	Alessia Milani},
	title        = {Twins: {BFT} Systems Made Robust},
	booktitle    = {25th International Conference on Principles of Distributed Systems,
	{OPODIS} 2021, Strasbourg, France, December 13-15, 2021},
	series       = {LIPIcs},
	volume       = {217},
	pages        = {7:1--7:29},
	publisher    = {Schloss Dagstuhl - Leibniz-Zentrum f{\"u}r Informatik},
	year         = {2021},
}

@article{DBLP:journals/corr/abs-2209-13304,
	author       = {Martina Camaioni and
	Rachid Guerraoui and
	Matteo Monti and
	Manuel Vidigueira},
	title        = {Oracular {Byzantine} Reliable Broadcast},
	journal      = {CoRR},
	volume       = {abs/2209.13304},
	year         = {2022},
}

@inproceedings{DBLP:conf/opodis/AuvolatRT19,
	author       = {Alex Auvolat and
	Michel Raynal and
	Fran{\c{c}}ois Ta{\"i}ani},
	editor       = {Pascal Felber and
	Roy Friedman and
	Seth Gilbert and
	Avery Miller},
	title        = {{Byzantine}-Tolerant Set-Constrained Delivery Broadcast},
	booktitle    = {23rd International Conference on Principles of Distributed Systems,
	{OPODIS} 2019, Neuch{\^{a}}tel, Switzerland, December 17-19, 2019},
	series       = {LIPIcs},
	volume       = {153},
	pages        = {6:1--6:23},
	publisher    = {Schloss Dagstuhl - Leibniz-Zentrum f{\"u}r Informatik},
	year         = {2019},
}

@article{DBLP:journals/tcs/AlbouyFRT23,
	author       = {Timoth{\'e} Albouy and
	Davide Frey and
	Michel Raynal and
	Fran{\c{c}}ois Ta{\"i}ani},
	title        = {Asynchronous {Byzantine} reliable broadcast with a message adversary},
	journal      = {Theor. Comput. Sci.},
	volume       = {978},
	pages        = {114110},
	year         = {2023},
}

@inproceedings{DBLP:conf/usenix/JansenNW22,
	author       = {Rob Jansen and
	James Newsome and
	Ryan Wails},
	editor       = {Jiri Schindler and
	Noa Zilberman},
	title        = {Co-opting Linux Processes for High-Performance Network Simulation},
	booktitle    = {Proceedings of the 2022 {USENIX} Annual Technical Conference, {USENIX}
	{ATC} 2022, Carlsbad, CA, USA, July 11-13, 2022},
	pages        = {327--350},
	publisher    = {{USENIX} Association},
	year         = {2022},
}

@inproceedings{DBLP:conf/prdc/BergerTR23,
	author       = {Christian Berger and
	Sadok Ben Toumia and
	Hans P. Reiser},
	title        = {Scalable Performance Evaluation of {Byzantine} Fault-Tolerant Systems Using Network Simulation},
	booktitle    = {28th {IEEE} Pacific Rim International Symposium on Dependable Computing,
	{PRDC} 2023, Singapore, October 24-27, 2023},
	pages        = {180--190},
	publisher    = {{IEEE}},
	year         = {2023},
}

@inproceedings{DBLP:conf/opodis/AlbouyFGHRSTZ24,
	author       = {Timoth{\'e} Albouy and
	Davide Frey and
	Ran Gelles and
	Carmit Hazay and
	Michel Raynal and
	Elad Michael Schiller and
	Fran{\c{c}}ois Ta{\"i}ani and
	Vassilis Zikas},
	editor       = {Silvia Bonomi and
	Letterio Galletta and
	Etienne Rivi{\`{e}}re and
	Valerio Schiavoni},
	title        = {Near-Optimal Communication {Byzantine} Reliable Broadcast Under a Message Adversary},
	booktitle    = {28th International Conference on Principles of Distributed Systems,
	{OPODIS} 2024, Lucca, Italy, December 11-13, 2024},
	series       = {LIPIcs},
	volume       = {324},
	pages        = {14:1--14:29},
	publisher    = {Schloss Dagstuhl - Leibniz-Zentrum f{\"u}r Informatik},
	year         = {2024},
}

@inproceedings{DBLP:conf/fmbc/NetoO25,
	author       = {Jo{\~a}o Miguel Louro Neto and
	Burcu Kulahcioglu Ozkan},
	editor       = {Diego Marmsoler and
	Meng Xu},
	title        = {A Benchmark Framework for {Byzantine} Fault Tolerance Testing Algorithms (Tool Paper)},
	booktitle    = {6th International Workshop on Formal Methods for Blockchains, {FMBC}
	2025, Hamilton, Canada, May 4, 2025},
	series       = {OASIcs},
	pages        = {13:1--13:11},
	publisher    = {Schloss Dagstuhl - Leibniz-Zentrum f{\"u}r Informatik},
	year         = {2025},
}

@article{DBLP:journals/iandc/Bracha87,
	author       = {Gabriel Bracha},
	title        = {Asynchronous {Byzantine} Agreement Protocols},
	journal      = {Inf. Comput.},
	volume       = {75},
	number       = {2},
	pages        = {130--143},
	year         = {1987},
}

@misc{klauspost-reedsolomon,
	author       = {Post, Klaus and others},
	title        = {klauspost/reedsolomon: v1.13.3},
	year         = {2026},
	version      = {v1.13.3},
	url          = {https://github.com/klauspost/reedsolomon},
	urldate      = {2026-04-22},
}

@misc{gnark-crypto-v0.20.1,
	author       = {Gautam Botrel and
	Thomas Piellard and
	Youssef El Housni and
	Arya Tabaie and
	Gus Gutoski and
	Ivo Kubjas and
	Yao J. Galteland},
	title        = {Consensys/gnark-crypto: v0.20.1},
	month        = mar,
	year         = 2026,
	publisher    = {Zenodo},
	version      = {v0.20.1},
}

@misc{shadow_guide,
	author       = {Rob Jansen and others},
	title        = {The {Shadow} Simulator},
	url          = {https://shadow.github.io/docs/guide/},
}

@inproceedings{DBLP:conf/nsdi/SinghDMDR08,
	author       = {Atul Singh and
	Tathagata Das and
	Petros Maniatis and
	Peter Druschel and
	Timothy Roscoe},
	editor       = {Jon Crowcroft and
	Michael Dahlin},
	title        = {{BFT} Protocols Under Fire},
	booktitle    = {5th {USENIX} Symposium on Networked Systems Design {\&} Implementation,
	{NSDI} 2008, April 16-18, 2008, San Francisco, CA, USA, Proceedings},
	pages        = {189--204},
	publisher    = {{USENIX} Association},
	year         = {2008},
}

@article{DBLP:journals/fac/BergerTR24,
	author       = {Christian Berger and
	Sadok Ben Toumia and
	Hans P. Reiser},
	title        = {Exploring Scalability of {BFT} Blockchain Protocols through Network Simulations},
	journal      = {Formal Aspects Comput.},
	volume       = {36},
	number       = {4},
	pages        = {24:1--24:29},
	year         = {2024},
}

@article{DBLP:journals/pacmpl/WinterBGGO23,
	author       = {Levin N. Winter and
	Florena Buse and
	Daan de Graaf and
	Klaus von Gleissenthall and
	Burcu Kulahcioglu Ozkan},
	title        = {Randomized Testing of {Byzantine} Fault Tolerant Algorithms},
	journal      = {Proc. {ACM} Program. Lang.},
	volume       = {7},
	number       = {{OOPSLA1}},
	pages        = {757--788},
	year         = {2023},
}

@article{DBLP:journals/cn/WangZWWH24,
	author       = {Jitao Wang and
	Bo Zhang and
	Kai Wang and
	Yuzhou Wang and
	Weili Han},
	title        = {BFTDiagnosis: An automated security testing framework with malicious behavior injection for {BFT} protocols},
	journal      = {Comput. Networks},
	volume       = {249},
	pages        = {110404},
	year         = {2024},
}

@inproceedings{DBLP:conf/podc/WanM0SX23,
	author       = {Jun Wan and
	Atsuki Momose and
	Ling Ren and
	Elaine Shi and
	Zhuolun Xiang},
	editor       = {Rotem Oshman and
	Alexandre Nolin and
	Magn{\'{u}}s M. Halld{\'{o}}rsson and
	Alkida Balliu},
	title        = {On the Amortized Communication Complexity of {Byzantine} Broadcast},
	booktitle    = {Proceedings of the 2023 {ACM} Symposium on Principles of Distributed Computing, {PODC} 2023, Orlando, FL, USA, June 19-23, 2023},
	pages        = {253--261},
	publisher    = {{ACM}},
	year         = {2023},
}

@inproceedings{DBLP:conf/ccs/MillerXCSS16,
	author       = {Andrew Miller and
	Yu Xia and
	Kyle Croman and
	Elaine Shi and
	Dawn Song},
	editor       = {Edgar R. Weippl and
	Stefan Katzenbeisser and
	Christopher Kruegel and
	Andrew C. Myers and
	Shai Halevi},
	title        = {The Honey Badger of {BFT} Protocols},
	booktitle    = {Proceedings of the 2016 {ACM} {SIGSAC} Conference on Computer and Communications Security, Vienna, Austria, October 24-28, 2016},
	pages        = {31--42},
	publisher    = {{ACM}},
	year         = {2016},
}

@inproceedings{DBLP:conf/eurosys/DanezisKSS22,
	author       = {George Danezis and
	Lefteris Kokoris{-}Kogias and
	Alberto Sonnino and
	Alexander Spiegelman},
	editor       = {Y{\'{e}}rom{-}David Bromberg and
	Anne{-}Marie Kermarrec and
	Christos Kozyrakis},
	title        = {Narwhal and Tusk: a DAG-based mempool and efficient {BFT} consensus},
	booktitle    = {EuroSys '22: Seventeenth European Conference on Computer Systems, Rennes, France, April 5 - 8, 2022},
	pages        = {34--50},
	publisher    = {{ACM}},
	year         = {2022},
}

@article{DBLP:journals/corr/abs-2104-03673,
  author       = {Silvia Bonomi and
                  J{\'{e}}r{\'{e}}mie Decouchant and
                  Giovanni Farina and
                  Vincent Rahli and
                  S{\'{e}}bastien Tixeuil},
  title        = {Practical {Byzantine} Reliable Broadcast on Partially Connected Networks},
  journal      = {CoRR},
  volume       = {abs/2104.03673},
  year         = {2021},
}

@article{DBLP:journals/eatcs/AuvolatFRT20,
	author       = {Alex Auvolat and
	Davide Frey and
	Michel Raynal and
	Fran{\c{c}}ois Ta{\"i}ani},
	title        = {Money Transfer Made Simple: a Specification, a Generic Algorithm, and its Proof},
	journal      = {Bull. {EATCS}},
	volume       = {132},
	year         = {2020},
}

@inproceedings{DBLP:conf/dsn/CollinsGKKMPPST20,
  author       = {Daniel Collins and
                  Rachid Guerraoui and
                  Jovan Komatovic and
                  Petr Kuznetsov and
                  Matteo Monti and
                  Matej Pavlovic and
                  Yvonne{-}Anne Pignolet and
                  Dragos{-}Adrian Seredinschi and
                  Andrei Tonkikh and
                  Athanasios Xygkis},
  title        = {Online Payments by Merely Broadcasting Messages},
  booktitle    = {50th Annual {IEEE/IFIP} International Conference on Dependable Systems
                  and Networks, {DSN} 2020, Valencia, Spain, June 29 - July 2, 2020},
  pages        = {26--38},
  publisher    = {{IEEE}},
  year         = {2020}
}

@article{DBLP:journals/corr/abs-2007-14990,
	author       = {Yingjian Wu and
	Haochen Pan and
	Saptaparni Kumar and
	Lewis Tseng},
	title        = {Reliable Broadcast in Practical Networks: Algorithm and Evaluation},
	journal      = {CoRR},
	volume       = {abs/2007.14990},
	year         = {2020},
}

@article{DBLP:journals/tcs/AuvolatFRT21,
	author       = {Alex Auvolat and
	Davide Frey and
	Michel Raynal and
	Fran{\c{c}}ois Ta{\"i}ani},
	title        = {{Byzantine}-tolerant causal broadcast},
	journal      = {Theor. Comput. Sci.},
	volume       = {885},
	pages        = {55--68},
	year         = {2021},
}

@inproceedings{DBLP:conf/focs/Dolev81,
	author       = {Danny Dolev},
	title        = {Unanimity in an Unknown and Unreliable Environment},
	booktitle    = {22nd Annual Symposium on Foundations of Computer Science, Nashville,
	Tennessee, USA, 28-30 October 1981},
	pages        = {159--168},
	publisher    = {{IEEE} Computer Society},
	year         = {1981},
}

@phdthesis{DBLP:phd/hal/Albouy24,
  author       = {Timoth{\'{e}} Albouy},
  title        = {Foundations of Reliable Cooperation under Asynchrony, Byzantine Faults, and Message Adversaries},
  school       = {University of Rennes, France},
  year         = {2024},
  url          = {https://tel.archives-ouvertes.fr/tel-04764046},
}

\end{document}